\documentclass[prd,amsfonts,noshowpacs,nofootinbib]{revtex4}
\usepackage{amsmath,graphicx}

\begin{document}
\title{On the full quantum trispectrum  in multi-field
DBI inflation}
\author{Shuntaro Mizuno\footnote{shuntaro.mizuno@nottingham.ac.uk}$\sharp$}
\author{Frederico Arroja\footnote{arrojaf@yukawa.kyoto-u.ac.jp}$\flat$}
\author{Kazuya Koyama\footnote{Kazuya.Koyama@port.ac.uk}$\natural$}
\affiliation{
$\sharp$School of Physics and Astronomy, University of Nottingham, University Park, Nottingham NG7 2RD, UK;
Research Center for the Early Universe (RESCEU), Graduate School of Science, The University of Tokyo, Tokyo 113-0033, Japan.
\\
$\flat$Yukawa Institute for Theoretical Physics, Kyoto University, Kyoto 606-8502, Japan.
\\
$\natural$Institute of Cosmology and Gravitation, University of Portsmouth, Portsmouth PO1 3FX, UK.
}

\date{\today}
\begin{abstract}
We compute the leading order connected four-point
function of the primordial curvature perturbation
coming from the four-point function of the fields
in multi-field DBI inflation models.
We confirm that the consistency relations in the squeezed limit
and in the counter-collinear limit hold as in single field models
thanks to special properties of the DBI action.
We also study the momentum dependence of the trispectra
coming from the adiabatic, mixed and purely entropic
contributions separately and we find that they have different momentum
dependence. This means that if the amount of the transfer from
the entropy perturbations to the curvature perturbation is significantly large,
the trispectrum can distinguish multi-field DBI inflation
models from single field DBI inflation models.
A large amount of transfer $T_{\mathcal{RS}} \gg 1 $ suppresses the tensor to scalar ratio $r \propto T_{\mathcal{RS}}^{-2}$ and the amplitude of the bispectrum $f_{NL}^{equi} \propto T_{\mathcal{RS}}^{-2}$ and so it can ease the severe observational constraints on the DBI inflation model based on string theory. On the other hand, it enhances the amplitude of the trispectrum
$\tau_{NL}^{equi} \propto T_{\mathcal{RS}}^2 f_{NL}^{equi \;2}$
for a given amplitude of the bispectrum.
\end{abstract}

\maketitle

\section{Introduction}

Precise measurements of the cosmic microwave background (CMB)
anisotropies such as those obtained by the WMAP satellite \cite{WMAP} provide
valuable information on the very early universe.
Any theoretical model that attempts to explain
the evolution of the universe before
the big bang nucleosynthesis will also
have to explain the observed CMB anisotropies.
Even though these anisotropies are almost
Gaussian, a small amount of non-Gaussianity is still allowed by the data
\cite{Yadav:2007yy,Komatsu:2008hk,Smith:2009jr}
and future experiments such as PLANCK \cite{Planck}
might detect this small amount of the
primordial non-Gaussianity.
Even though
the simplest slow-roll single field inflation models
predict that the non-Gaussianity of
the fluctuations should be very difficult
to be detected \cite{Maldacena:2002vr},
recently, theoretical models which can produce
sizeable non-Gaussianity has been
extensively studied by many authors
\cite{Linde:1996gt,Bartolo:2001cw,Bernardeau:2002jy,Bernardeau:2002jf,Dvali:2003em,Creminelli:2003iq,Alishahiha:2004eh,Gruzinov:2004jx,Bartolo:2004if,Enqvist:2004ey,Seery:2005wm,Seery:2005gb,Jokinen:2005by,Lyth:2005qk,Salem:2005nd,Seery:2006js,
Sasaki:2006kq,Malik:2006pm,Barnaby:2006cq,Alabidi:2006wa,Chen:2006nt,Huang:2006eh,Chen:2006xjb,Alabidi:2006hg,Seery:2006vu,Byrnes:2006vq,Suyama:2007bg,Arroja:2008ga,Arroja:2008yy,Langlois:2008wt,Langlois:2008qf,Seery:2008ax,
Sasaki:2008uc,Byrnes:2008wi,Byrnes:2008zy,Dutta:2008if,Naruko:2008sq,Suyama:2008nt,Gao:2008dt,Cogollo:2008bi,Rodriguez:2008hy,Ichikawa:2008iq,Byrnes:2008zz,Li:2008fma,Langlois:2008vk,Hikage:2008sk,
Kawasaki:2008sn,Huang:2008zj,Gao:2009gd,Cai:2009hw,Langlois:2009ej,Gao:2009bx,Huang:2009xa,Huang:2009vk,Khoury:2008wj,Arroja:2009pd,Chen:2009bc,Mizuno:2009cv,Byrnes:2009qy,Gao:2009at}.

Among them, the Dirac-Born-Infeld (DBI) inflation model
can produce large non-Gaussianity by the fact that
the sound speed of the perturbations can be much smaller
than one due to the non-trivial form of the kinetic term
\cite{ArmendarizPicon:1999rj,Garriga:1999vw}. DBI inflation models are also
well motivated by string theory
\cite{Silverstein:2003hf,Alishahiha:2004eh,Chen:2004gc,Chen:2005ad,Chen:2005fe,Chen:2006nt}.
In this model, the inflaton fields are identified
with the positions of a moving D3 brane in the six-dimensional internal space. The dynamics of the D3 brane
is described by the DBI action.
However, recently it has been pointed out
that single field DBI inflation driven by
a mobile D3 brane with large non-Gaussianity might
contradict with the current WMAP data.
For current and stringent observational constraints and consequences of DBI-inflation see
\cite{Kecskemeti:2006cg,Lidsey:2006ia,Baumann:2006cd,Bean:2007hc,Lidsey:2007gq,Peiris:2007gz,Kobayashi:2007hm,Gmeiner:2007uw,Lorenz:2007ze,Bean:2007eh,Bird:2009pq}.

One way to avoid these constraints is to consider
multi-field DBI models \cite{Langlois:2008wt}.
Since the position of the brane in each compact direction is
described by a scalar field, DBI inflation is naturally a
multi-field inflationary model \cite{Easson:2007dh}.
As first pointed out by \cite{Starobinsky:1994mh},
in multi-field inflation models,
the curvature perturbation is modified
on large scales due to the entropy perturbation.
Even though there are some works considering
multi-field inflationary models with non-canonical
kinetic terms
\cite{Huang:2007hh,Langlois:2008mn,Gao:2008dt,RenauxPetel:2008gi},
the consistent analysis for the entropy modes
in the multi-field
DBI inflation model has started only very recently
\cite{Langlois:2008wt,Langlois:2008qf,Arroja:2008yy}.

In \cite{Langlois:2008wt,Langlois:2008qf,Arroja:2008yy},
the three-point functions in the small sound speed limit and
at leading order in the slow-roll expansion were obtained
and it was shown that in addition to the purely adiabatic
three-point function, there exists a mixed component
$\langle
Q_{\sigma}(\mathbf{k_1})Q_{s}(\mathbf{k_2})Q_{s}(\mathbf{k_3})
\rangle$ where $Q_\sigma$ and $Q_s$ are the adiabatic
and the entropy perturbations, respectively.
Since the momentum dependence of the three-point function
from the adiabatic modes was shown to be identical
with the mixed component, the shape of the bispectrum of
the curvature perturbations remains the same as in the single-field
case, while the amplitude is affected by the entropy
perturbation.

Even though previous works on the non-Gaussianity
in the multi-field DBI inflation model are
mainly limited to the bispectrum,
it is expected that the cosmic microwave background (CMB)
trispectrum also provides strong constraints
on early universe models.
At the moment, the constraints are rather weak given by
$|\tau_{NL}| < 10^8$ \cite{Boubekeur:2005fj,Alabidi:2005qi},
where $\tau_{NL}$ denotes the size of the trispectrum.
However, PLANCK will tighten the constraints
significantly reaching
$|\tau_{NL}| \sim 560$ \cite{Kogo:2006kh}.
Although these estimations are obtained assuming local type non-Gaussianity,
we expect similar constraints can be obtained for non-Gaussianity from
DBI inflation. It is also worth noting that
the analysis in the single field DBI inflation model
shows that the trispectrum from the contact interaction
diagram is enhanced in the small sound speed limit as
$\tau_{NL} \sim 1/c_s^4$
\cite{Huang:2006eh}. Even though the contribution
from the scalar exchange interaction was overlooked
in this analysis, recently, it has been confirmed
that the same scaling holds for the trispectrum
from this interaction \cite{Arroja:2009pd,Chen:2009bc}.
As in the bispectrum case \cite{Babich:2004gb},
the observational constraints depend
on the shape of the wave vectors' configuration.
Therefore, it is important
to calculate the full shape dependence of the trispectrum
in the multi-field DBI inflation model.
For the details of the observations
of the CMB trispectrum, see
\cite{Hu:2001fa,Okamoto:2002ik,Creminelli:2006gc}.

For this purpose, recently
we have calculated the quantum trispectrum
from the contact interaction
and obtained the simple understanding of the origin
of the interaction terms of the multi-field DBI inflation
\cite{Mizuno:2009cv}. In this paper,
as a natural continuation,  we will calculate
the scalar exchange trispectrum
and obtain the complete theoretical prediction
for the quantum trispectrum of multi-field DBI inflation.
For a related work, where only the purely entropic
component of the quantum scalar exchange trispectrum
was calculated, see
version three of \cite{Gao:2009gd}.

The structure of this paper is as follows.
In the next section,
we shall introduce the model and some basic notation.
In section \ref{sec:Contact} we summarise the results
obtained in \cite{Mizuno:2009cv} for the
contact interaction trispectrum.
After calculating the scalar exchange trispectrum
in section \ref{sec:FOURPOINTFUNCTION_exchange},
the momentum dependence of the trispectra
coming from the adiabatic, mixed, and purely entropic contributions
are studied separately
in section \ref{sec:shape}. The observational
constraints on the multi-field DBI inflation model are discussed in
section \ref{sec:OBSERVATIONS}. Section \ref{sec:CONCLUSIONS} is devoted to the conclusions.

\section{\label{sec:MODEL}The model}

In this section, we will introduce the multi-field Dirac-Born-Infeld (DBI) inflationary model. We will present the background evolution equations and define some basic notation.

The multi-field DBI inflation model is described by the following action
\cite{Leigh:1989jq}
\begin{eqnarray}
&&S=\frac{1}{2}\int
d^4x\sqrt{-g}\left[R+2\tilde{P}(\tilde{X},\phi^I)\right]\,,
\nonumber\\
&&\tilde{P} (\tilde{X},\phi^I) = -\frac{1}{f(\phi^I)}
\left( \sqrt{1-2f(\phi^I) \tilde{X}}-1 \right)-V(\phi^I)\,,
\label{action}
\end{eqnarray}
where we have set $8 \pi G =1$, $R$  is the Ricci scalar,
$\phi^I$ are the scalar fields $(I=1,2,...,N_\phi)$,
$f(\phi^I)$ and $V(\phi^I)$ are functions of
the scalar fields determined by string theory
configurations
and $\tilde{X}$ is defined in terms of the determinant
${\cal D} \equiv \mbox{det} (\delta^\mu_\nu + f G_{IJ}
\partial^\mu \phi^I \partial_\nu \phi^J)$ as
$\tilde{X} = (1- {\cal D})/ (2f).$
Here $G_{IJ}$ is the metric in the field space.
We assume that $\tilde{P}$ is a well behaved function
of $\phi^I$ and $\tilde{X}$.
It is also shown that
$\tilde{X}$ is related to the kinetic term of the scalar
fields as \cite{Langlois:2008wt,Langlois:2008qf}
\begin{eqnarray}
\tilde{X} &=&  G_{IJ} X^{IJ} -2f  X_I ^{\;[I} X_J ^{\;J]}
+ 4f^2 X_I ^{\;[I} X_J ^{\;J} X_K ^{\;K]}
-8f^3  X_I ^{\;[I} X_J ^{\;J} X_K ^{\;K}
X_L ^{\;L]}\,,\\
X^{IJ} &\equiv& -\frac12 g^{\mu\nu} \partial_{\mu} \phi^I
\partial_\nu \phi^J\,,\;\;\;\;
X_{I}^{\;J} = G_{IK} X^{KJ}\,,
\end{eqnarray}
where the brackets denote antisymmetrization.
It is worth noting that even though
$\tilde{X} = X \,(\,= G_{IJ} X^{IJ})$ in the homogeneous
background, this does not hold if we take into account
the inhomogeneous components.

In the background, we are interested in flat, homogeneous and
isotropic Friedman-Robertson-Walker (FRW) universes described by the
line element
\begin{equation}
ds^2=-dt^2+a^2(t)\delta_{ij}dx^idx^j, \label{FRW}
\end{equation}
where $a(t)$ is the scale factor.
 The Friedman equation and the
continuity equation read
\begin{equation}
3H^2=E_0, \label{EinsteinEq}
\end{equation}
\begin{equation}
\dot{E}_0=-3H\left(E_0+\tilde{P}_0 \right), \label{continuity}
\end{equation}
where the Hubble rate is $H=\dot{a}/a$,
a dot denotes derivative with respect to cosmic time
$t$, $E_0$ is the total energy
of the fields which is given by
\begin{equation}
E_0=2X_0^{IJ}\tilde{P}_{0,X^{IJ}}-\tilde{P}_0 ,\label{energy}
\end{equation}
and  the subscript $0$ denotes that the quantity is evaluated
in the background.

For this model the
speed of propagation of scalar perturbations
(``speed of sound"), $c_s$, is given by
\begin{equation}
c_s^2 \equiv \left(\frac{\tilde{P}_{,\tilde{X}}}
{\tilde{P}_{,\tilde{X}}+ 2\tilde{X} \tilde{P}_{,\tilde{X}
\tilde{X}}}\right)_0\,.\label{DBIcs}
\end{equation}

Since we are interested in the inflationary background,
we assume that the form of $f(\phi^I)$ and $V(\phi^I)$
are chosen so that inflation is realized at least
for $60$ e-foldings. In order to characterize
this background,
we define the slow variation parameters, analogues of the
slow-roll parameters, as:
\begin{equation}
\epsilon=-\frac{\dot{H}}{H^2}=
\frac{X_0}
{H^2 c_{s}}, \quad
\iota=\frac{\dot{\epsilon}}{\epsilon H}, \quad
\chi=\frac{\dot{c_s}}{c_sH}.
\label{slow_roll_parameters}
\end{equation}
We should note that these slow variation parameters are more
general than the usual slow-roll parameters and that the smallness
of these parameters does not imply that the field is rolling
slowly.
We assume that the rate of change of the speed of sound is
small (as described by $\chi$) but $c_s$ is otherwise free to change
between zero and one.

We shall consider perturbations on this FRW background.
We decompose the scalar field $\phi^I$
into the background value $\phi_0^I$ and
perturbation $Q^I$ in the flat gauge as,
\begin{eqnarray}
\phi^I (x,t) = \phi^I_0 (t) + Q^I (x,t)\,.
\end{eqnarray}

Furthermore, as was done in \cite{Gordon:2000hv},
we decompose the perturbations into
instantaneous adiabatic and entropy perturbations,
where the adiabatic direction corresponds to
the direction of the background fields' evolution
while the entropy directions are orthogonal to this.
We introduce an orthogonal basis
${e_{n}^I}$, with $n=1,2,...,N_\phi$, in the field space
so that the orthonormality condition are given by
\cite{Arroja:2008yy}
\begin{equation}
e_{n}^I e_{m I} =
\frac{1}{c_s}
\delta_{mn}-
\frac{1-c_s^2}{c_s} \delta_{m1} \delta_{n1}\,,
\label{orthonormality_cond}
\end{equation}
where the adiabatic basis is defined as
\begin{equation}
e^I_1 = \sqrt{\frac{c_s}{2X_0}} \dot{\phi}_0 ^I\,.
\label{cond_ad_base}
\end{equation}
Notice that the length of the basis vector $e^I _{0}$
is $c_s$ and that of the other basis vectors is
$1/c_s$.
If we consider the two-field case ($I=1,2$),
the field perturbations are decomposed
into the adiabatic field $Q_\sigma$ and the entropy field
$Q_s$ as
\begin{eqnarray}
Q^I = Q_\sigma e^I_1 + Q_s e^I_2\,.
\label{int_qsigma_qs}
\end{eqnarray}
Hereafter, for simplicity, we will concentrate on
the two-field case although the extension to
more fields is straightforward.

For the single field model, since the comoving curvature
perturbation $\mathcal{R}$ is given by
\begin{eqnarray}
\mathcal{R} = \bigl(\frac{\sqrt{c_s}H}{\sqrt{2 X_0}}\bigr)_*
Q_{\sigma *}\,,
\end{eqnarray}
the power spectrum of the primordial quantum
fluctuation was given by \cite{Garriga:1999vw}
\begin{equation}
{\cal P}_\mathcal{R_*}(k)=\frac{1}{36\pi^2}\frac{E_0^2}
{E_0+\tilde{P}_0}=\frac{1}{8\pi^2}\frac{H^2}{c_s\epsilon},
\label{PowerSpectrum}
\end{equation}
where it should be evaluated at the time of horizon crossing
${c_s}_*k=a_*H_*$. The spectral index is
\begin{equation}
n_\mathcal{R_*}-1=\frac{d\ln {\mathcal{P}}_\mathcal{R_*}(k)}
{d\ln k}=-2\epsilon-
\iota-\chi.
\label{SpectralIndex}
\end{equation}
However, as explained in the following sections,
this will not be the case for multi-field models.

\section{\label{sec:Contact}The Trispectrum from the contact interaction}

In this and next sections, we will derive the four-point functions of
the field perturbations
at leading order in the slow-roll expansion
and in the small sound speed limit.
There are two important tree-level
diagrams for the trispectrum. One is a diagram
where the interaction occurs at a point,
i.e. a contact interaction diagram and
the other  is a diagram where
a scalar mode is exchanged.
In the single field DBI inflation model, it was recently shown by \cite{Arroja:2009pd,Chen:2009bc} that both contributions are comparable.
Because it is expected that this is also true
for the two-field DBI inflation model,
it is necessary to consider both contributions.
In this section, we summarize the calculation of
the four-point functions of the field perturbations
at horizon crossing coming from
the contact interaction diagram \cite{Mizuno:2009cv}. In the next section, we shall calculate the scalar exchange four-point function.

Using the approximations mentioned above and following
the ADM formalism
\cite{Arnowitt:1960es,Maldacena:2002vr,Seery:2005wm,Seery:2005gb},
the action up to fourth order can be calculated as \cite{Mizuno:2009cv}
\begin{eqnarray}
&&S^{(main)}_{(2)}= \frac12 \int dt d^3x
\frac{a^3}{c_s^2}
\biggl[\dot{Q}_\sigma^2+
\dot{Q}_s^2-\frac{c_s^2}{a^2}  \bigg(
 \partial_i Q_\sigma \partial^i Q_\sigma
+ \partial_i Q_s \partial^i Q_s\bigg) \biggr]\,,
\label{two_field_lead_dbi_sec}\\
&&S^{(main)}_{(3)}= \frac12 \int dt d^3 x
\frac{a^3}{\sqrt{2X_0 c_s^7}} \biggl[
\dot{Q}_\sigma^3
+
\dot{Q}_\sigma \dot{Q}_s^2
+ \frac{c_s^2}{a^2} \bigg(
\left( \partial_i Q_s \partial^i Q_s
- \partial_i Q_\sigma \partial^i Q_\sigma
\right) \dot{Q}_\sigma
-2
\left(\partial_i Q_{\sigma} \partial^i Q_s \right)
\dot{Q}_s \bigg)
\biggr]\,,\label{two_field_lead_dbi_thir}\\
&&S^{(main)}_{(4)} = \frac{1}{16} \int dx^3 dt
\frac{a^3}{c_s^5 X_0}
\Biggl[
5 \dot{Q}_\sigma^4 +
6 \dot{Q}_\sigma^2 \dot{Q}_s^2
+\dot{Q}_s^4 -\frac{2 c_s^2}{a^2}
\bigg(
3 \dot{Q}_\sigma^2 \partial_i Q_\sigma \partial^i Q_\sigma
- \dot{Q}_\sigma^2 \partial_i Q_s \partial^i Q_s
+4 \dot{Q}_\sigma \dot{Q}_s
\partial_i Q_\sigma \partial^i Q_s
\nonumber\\
&&\qquad\qquad\qquad\qquad\qquad\qquad\qquad\qquad\qquad\qquad\qquad\qquad\quad
+ \dot{Q}_s^2  \partial_i Q_\sigma \partial^i Q_\sigma
+
\dot{Q}_s^2  \partial_i Q_s \partial^i Q_s
\bigg)\nonumber\\
&&\qquad\qquad\qquad\qquad\qquad\qquad
+ \frac{c_s^4}{a^4}  \bigg(
\left(\partial_i Q_\sigma \partial^i Q_\sigma\right)^2
 -2 \left(\partial_i Q_\sigma \partial^i Q_\sigma\right)
\left(\partial_j Q_s \partial^j Q_s\right)
+4 \left(\partial_i Q_\sigma \partial^i Q_s\right)^2
+\left(\partial_i Q_s \partial^i Q_s\right)^2
\bigg)\Biggr]\,.\nonumber\\
\label{two_field_lead_dbi_four}
\end{eqnarray}
It is worth noting that Eqs.~(\ref{two_field_lead_dbi_sec}),
(\ref{two_field_lead_dbi_thir}) and
(\ref{two_field_lead_dbi_four})
can also be derived using a simpler method based on a Lorentz boost
from the frame where the brane is at rest to the frame where the
brane is moving \cite{Mizuno:2009cv}.

Based on these actions,
we can obtain the third- and fourth-order interaction
Hamiltonian densities as \cite{Mizuno:2009cv}
\begin{eqnarray}
{{\cal H}}_{(3)} ^{int} &=&
\frac{-a^3}{2 \sqrt{2 X_0 c_s^7}}
\bigg[\dot{Q}_\sigma ^3 + \dot{Q}_\sigma \dot{Q}_s^2-
\frac{c_s^2}{a^2} \left\{ \dot{Q}_\sigma
\left(\partial_i Q_\sigma \partial^i Q_\sigma
- \partial_i Q_s \partial^i Q_s\right)
+ 2 \dot{Q}_s \partial_i Q_\sigma \partial^i Q_s\right\}
\bigg]\,,\label{two_field_lead_dbi_hamil_third}\\
{{\cal H}}_{(4)} ^{int} &=& \frac{a^3}{4 X_0 c_s^5}
\biggl[ \dot{Q}_\sigma^4
+ \dot{Q}_\sigma^2 \dot{Q}_s^2
+ \frac{c_s^2}{a^2}
\left\{\left(\partial_i Q_s \partial^i Q_s
\right) \dot{Q}_\sigma^2
+  \left(\partial_i Q_s \partial^i Q_s
\right) \dot{Q}_s^2 \right\} \biggr]\,,
\label{two_field_lead_dbi_hamil_four}
\end{eqnarray}
where here, $Q_\sigma$ and $Q_s$ are the interaction picture fields.
One important remark is that
while the cubic part of  ${\cal{H}}^{int}$
is the opposite sign of the cubic Lagrangian density ${\cal{L}}^{int}$,
this is generally not true at fourth order.

From the second-order action,
we can solve for the perturbations and
quantize them according to the standard procedures
of quantum field theory:
\begin{equation}
Q_n(\eta,\mathbf{x})=\frac{1}{(2\pi)^3}\int
d^3\mathbf{k}\left[
u_n(\eta,\mathbf{k})a_n(\mathbf{k})+
u^*_n(\eta,-\mathbf{k})a^\dag_n(-\mathbf{k})\right]
e^{i\mathbf{k}\cdot\mathbf{x}}\,,
\end{equation}
where
$a_n(\mathbf{k})$ and $a^\dag_n(-\mathbf{k})$
are the annihilation
and creation operator respectively, which satisfy the usual
commutation relations:
\begin{eqnarray}
&&\left[a_n(\mathbf{k_1}),a^\dag_m(\mathbf{k_2})\right]
=(2\pi)^3\delta^{(3)}(\mathbf{k_1}-\mathbf{k_2})\delta_{nm}\,,
{\hspace{1cm}}
\left[a_n(\mathbf{k_1}),a_m(\mathbf{k_2})\right]
=\left[a_n^\dag(\mathbf{k_1}),a_m^\dag(\mathbf{k_2})\right]
=0\,.
\end{eqnarray}
At leading order, solutions for the mode functions are given by
\begin{equation}
u_n(\eta,\mathbf{k})
=A_n\frac{1}{k^{3/2}}\left(1+ikc_s\eta\right)e^{-ikc_s\eta}\,.
\end{equation}
The two point correlation function is then obtained as
\begin{eqnarray}
&&\langle
0|Q_n(\eta=0,\mathbf{k_1})Q_m(\eta=0,\mathbf{k_2})|0\rangle
=(2\pi)^3\delta^{(3)}(\mathbf{k_1}+\mathbf{k_2})
\mathcal{P}_{Q_n}\frac{2\pi^2}{k_1^3}\delta_{nm}\,,
\end{eqnarray}
where the power spectrum is defined as
\begin{equation}
\mathcal{P}_{Q_n}=\frac{|A_n|^2}{2\pi^2}\,, \quad
|A_\sigma|^2=|A_s|^2=\frac{H^2}{2c_{s}}\equiv N^2\,,
\end{equation}
and it should be evaluated at the time of the sound horizon crossing
${c_s}_* k_1=a_*H_*$.

In terms of these quantum operators,
the connected four-point correlation function
coming from the contact interaction diagram is given by
\cite{Weinberg:2005vy}
\begin{eqnarray}
&&\langle\Omega|Q_m(0,\mathbf{k_1})Q_n(0,\mathbf{k_2})
Q_p(0,\mathbf{k_3}) Q_q(0,\mathbf{k_4})|\Omega\rangle^{CI}
=\nonumber\\
&&{\hspace{4cm}}-i\int^0_{-\infty} d\eta
\langle 0
|\left[Q_m(0,\mathbf{k_1})Q_n(0,\mathbf{k_2})
Q_p(0,\mathbf{k_3}) Q_q(0,\mathbf{k_4}),
H_{(4)}^{int}(\eta) \right]|0\rangle\,,
\end{eqnarray}
where $Q_m$ on the r.h.s. of the equation are the interaction picture fields and $H_{(4)}^{int}$ is
the integral of the fourth-order Hamiltonian density
(\ref{two_field_lead_dbi_hamil_four}), that is,
$H_{(4)}^{int} = \int d^3 x {{\cal H}}_{(4)}^{int}$.

The purely adiabatic, purely entropic and
mixed components
are respectively given by
\begin{eqnarray}
&&
\label{4point_pureadiabatic}
\langle\Omega|Q_\sigma(0,\mathbf{k_1})Q_\sigma(0,\mathbf{k_2})
Q_\sigma(0,\mathbf{k_3})Q_\sigma(0,\mathbf{k_4})
|\Omega\rangle^{CI}=
(2 \pi)^3 \delta^{(3)} (\sum_{i=1}^4 \mathbf{k_i})
\frac{H^8}{2X_0 c_s^6}
\frac{1}{\Pi_{i=1}^4 k_i^3}
\left( -36 A_1 \right)\,,\\
&& \qquad\qquad\qquad\qquad A_1 = \frac{\Pi_{i=1}^4 k_i^2}{K^5}\,,
\quad K=\sum_{i=1} ^{4} k_i\,,\\
&&
\label{4point_pureentropy}
\langle\Omega|Q_s(0,\mathbf{k_1})Q_s(0,\mathbf{k_2})
Q_s(0,\mathbf{k_3})Q_s(0,\mathbf{k_4})
|\Omega\rangle^{CI}=
(2 \pi)^3 \delta^{(3)} (\sum_{i=1}^4 \mathbf{k_i})
\frac{H^8}{2X_0 c_s^6}
\frac{1}{\Pi_{i=1}^4 k_i^3}
\biggl(-
\frac{1}{8} A_2\biggr)\,,\\
&&\qquad\qquad\qquad\qquad
A_2 = \frac{k_1^2 k_2^2 (\mathbf{k_3}\cdot\mathbf{k_4})}
{K^3} \left( 1+ \frac{3(k_3 + k_4)}{K}
+ \frac{12 k_3 k_4}{K^2}\right)+{\rm perm.}\label{def_a2}\\
&&
\label{4point_mixed}
\langle\Omega|Q_\sigma(0,\mathbf{k_1})Q_\sigma(0,\mathbf{k_2})
Q_s(0,\mathbf{k_3})Q_s(0,\mathbf{k_4})
|\Omega\rangle^{CI}=
(2 \pi)^3 \delta^{(3)} (\sum_{i=1}^4 \mathbf{k_i})
\frac{H^8}{2X_0 c_s^6}
\frac{-1}{\Pi_{i=1}^4 k_i^3}
\biggl( 6 A_1 + \frac{1}{2} A_3 \biggr)\,,\\
&&
\qquad\qquad\qquad\qquad
A_3 = \frac{k_1^2 k_2^2 (\mathbf{k_3}\cdot\mathbf{k_4})}
{K^3} \left( 1+ \frac{3(k_3 + k_4)}{K}
+ \frac{12 k_3 k_4}{K^2}\right)\,,
\end{eqnarray}
where ``perm'' denotes twenty three permutations of $\{k_1,k_2,k_3,k_4\}$.

Following the analysis of
\cite{Wands:2002bn}, if
we describe the conversion of the entropy
perturbation into the curvature perturbation by a transfer
coefficient $T_{\mathcal{RS}}$,
the final comoving curvature perturbation
$\mathcal{R}$
is expressed in terms of the adiabatic and entropy
field perturbations as
\begin{eqnarray}
&&\mathcal{R}=\mathcal{A}_\sigma
Q_{\sigma *}+\mathcal{A}_sQ_{s*}\,,
\quad
\mathcal{A}_\sigma=
\left( \frac{\sqrt{c_s} H}{ \sqrt{2X_0}}\right)_*\,,
\quad \mathcal{A}_s= T_{\mathcal{RS}}
\left( \frac{\sqrt{c_s} H}{ \sqrt{2X_0}}\right)_*\,.
\label{curv_sum_ad_en}
\end{eqnarray}
This implies that the final comoving curvature perturbation power spectrum is
\begin{equation}
{\cal P}_\mathcal{R}=\left(1+T_{\mathcal{RS}}^2\right){\cal P}_\mathcal{R_*},
\end{equation}
where the horizon crossing power spectrum ${\cal P}_\mathcal{R_*}$ was
given by Eq. (\ref{PowerSpectrum}).

At leading order, the connected four-point function of $\mathcal{R}$
coming from the contact interaction diagram is given by
\footnote{There are other contributions to the connected four-point function of $\mathcal{R}$, for example those coming from the five- and six-point functions of the scalar fields that can be expressed in terms of the power spectrum using Wick's theorem. In this work, we ignore these contributions because we are only interested in the intrinsically quantum four-point function. These extra terms are generated by the non-linear evolution outside the horizon and can be important}
\begin{eqnarray}
\langle\mathcal{R}(\mathbf{k_1})\mathcal{R}(\mathbf{k_2})
\mathcal{R}(\mathbf{k_3}) \mathcal{R}(\mathbf{k_4})\rangle^{CI}
 &=& \mathcal{A}^4_\sigma
\langle
Q_\sigma(\mathbf{k_1})Q_\sigma(\mathbf{k_2})
Q_\sigma(\mathbf{k_3}) Q_\sigma(\mathbf{k_4})\rangle^{CI}
\nonumber\\
&&+ \mathcal{A}^2_\sigma \mathcal{A}^2_s
\big(\langle
Q_\sigma(\mathbf{k_1})Q_\sigma(\mathbf{k_2})
Q_s (\mathbf{k_3}) Q_s (\mathbf{k_4})\rangle^{CI}
+ {\rm 5\,perm.}\big)
\nonumber\\
&& + \mathcal{A}^4_s
\langle
Q_s (\mathbf{k_1})Q_s (\mathbf{k_2})
Q_s (\mathbf{k_3}) Q_s (\mathbf{k_4})\rangle^{CI}\,,
\label{4point_curv_cont}
\end{eqnarray}
where ``5 perm.'' denotes five permutations of
the four-momenta.
Using Eqs. (\ref{4point_pureadiabatic}), (\ref{4point_pureentropy}) and (\ref{4point_mixed}), it is possible to show a simple relation among the three different terms in the previous equation, as
\begin{eqnarray}
\langle
Q_\sigma(\mathbf{k_1})Q_\sigma(\mathbf{k_2})
Q_\sigma(\mathbf{k_3}) Q_\sigma(\mathbf{k_4})\rangle^{CI}
+
\langle
Q_s (\mathbf{k_1})Q_s (\mathbf{k_2})
Q_s (\mathbf{k_3}) Q_s (\mathbf{k_4})\rangle^{CI}
=
\langle
Q_\sigma(\mathbf{k_1})Q_\sigma(\mathbf{k_2})
Q_s (\mathbf{k_3}) Q_s (\mathbf{k_4})\rangle^{CI}
+ {\rm 5\,perm.}
\nonumber\\
\label{CI_simple_rel}
\end{eqnarray}
This relation is perhaps less surprising if one notes that the different fourth-order interactions present in Eq. (\ref{two_field_lead_dbi_hamil_four}) also obey a similar relation, i.e. the purely adiabatic plus the purely entropic interactions (vertices) are equal to the mixed vertices.

Using the previous results, the contact interaction four-point function of $\mathcal{R}$ can be written as
\begin{eqnarray}
\langle\mathcal{R}(\mathbf{k_1})\mathcal{R}(\mathbf{k_2})
\mathcal{R}(\mathbf{k_3}) \mathcal{R}(\mathbf{k_4})\rangle^{CI}
 &=& \mathcal{A}^4_\sigma\left(1+T_{\mathcal{RS}}^2\right)\bigg(
\langle
Q_\sigma(\mathbf{k_1})Q_\sigma(\mathbf{k_2})
Q_\sigma(\mathbf{k_3}) Q_\sigma(\mathbf{k_4})\rangle^{CI}
\nonumber\\
&&\qquad\qquad\qquad\quad
+T_{\mathcal{RS}}^2
\langle
Q_s (\mathbf{k_1})Q_s (\mathbf{k_2})
Q_s (\mathbf{k_3}) Q_s (\mathbf{k_4})\rangle^{CI}\bigg)\,.
\end{eqnarray}
One immediately sees that if the transfer coefficient is non-zero, one can in principle use the trispectrum to distinguish single field from multi-field DBI inflation.

\section{\label{sec:FOURPOINTFUNCTION_exchange}
The trispectrum from the scalar exchange interaction}

In this section, we will calculate
the four-point functions of the field perturbations
at horizon crossing coming from
a diagram where a scalar mode is exchanged.
Within the ``interaction picture" formalism \cite{Weinberg:2005vy},
to calculate the four-point function resulting
from a correlation established
via the exchange of a scalar mode,
one needs to evaluate the following time integrals
\begin{eqnarray}
&&\langle\Omega| Q_m(0,\mathbf{k_1})Q_n(0,\mathbf{k_2})Q_p(0,\mathbf{k_3})Q_q(0,\mathbf{k_4})|\Omega\rangle^{SE}
\nonumber\\
&&{\hspace{1cm}}= -\int_{-\infty}^0 d\eta\int_{-\infty}^\eta d\tilde\eta \langle
 0|\left[\left[Q_m(0,\mathbf{k_1})Q_n(0,\mathbf{k_2})
Q_p(0,\mathbf{k_3})Q_q(0,\mathbf{k_4}),H_{(3)}^{int}(\eta)\right],H^{int}_{(3)}(\tilde\eta)\right]|0\rangle, \label{SEintegral}
\end{eqnarray}
where $Q_m$ on the r.h.s. of the equation are the interaction picture fields and $H_{(3)}^{int}$ is
the integral of the third-order Hamiltonian density
(\ref{two_field_lead_dbi_hamil_third}), that is,
$H_{(3)}^{int} = \int d^3 x {{\cal H}}_{(3)}^{int}$.

The calculation of these time integrals is rather long and tedious. In Appendix \ref{diagramsapp}, we present an elegant and simple way of evaluating these contributions. This method is based on diagrams and simple rules that make
the origin and physical meaning of all the different terms in the following
expressions very clear. The purely adiabatic four-point function coming from the scalar exchange diagrams is identical to the single field one \cite{Arroja:2009pd}, and it is given by
\begin{eqnarray}
\langle\Omega|&&\!\!\!\!\!\!\!\!Q_\sigma(0,\mathbf{k_1})
Q_\sigma(0,\mathbf{k_2})Q_\sigma(0,\mathbf{k_3})
Q_\sigma(0,\mathbf{k_4})|\Omega\rangle^{SE}
=(2\pi)^3\delta^{(3)}(\mathbf{K})
\frac{H^4}{16 X_0 c_s^9(k_1k_2k_3k_4)^\frac{3}{2}}\times
\nonumber\\
&&\Bigg[
-9\bigg(\mathcal{F}_1(k_1,k_2,-k_{12},k_3,k_4,k_{12})-\mathcal{F}_1(-k_1,-k_2,-k_{12},k_3,k_4,k_{12})\bigg)
\nonumber\\&&
\,\,\,\,\,
- 3 c_s^2
\Bigg(
      (\mathbf{k_3}\cdot\mathbf{k_4})\bigg(\mathcal{F}_3(k_1,k_2,-k_{12},k_{12},k_3,k_4)-\mathcal{F}_3(-k_1,-k_2,-k_{12},k_{12},k_3,k_4)\bigg)
      \nonumber\\&&\quad\quad\quad\quad
      -2(\mathbf{k_{34}}\cdot\mathbf{k_4})\bigg(\mathcal{F}_3(k_1,k_2,-k_{12},k_3,k_4,k_{12})-\mathcal{F}_3(-k_1,-k_2,-k_{12},k_3,k_4,k_{12})\bigg)
      \nonumber\\&&\quad\quad\quad\quad
      +(\mathbf{k_1}\cdot\mathbf{k_2})\bigg(\mathcal{F}_4(-k_{12},k_1,k_2,k_3,k_4,k_{12})-\mathcal{F}_4(-k_{12},-k_1,-k_2,k_3,k_4,k_{12})\bigg)
      \nonumber\\&&\quad\quad\quad\quad
      -2(\mathbf{k_{12}}\cdot\mathbf{k_2})\bigg(\mathcal{F}_4(k_1,k_2,-k_{12},k_3,k_4,k_{12})-\mathcal{F}_4(-k_1,-k_2,-k_{12},k_3,k_4,k_{12})\bigg)
\Bigg)
\nonumber\\
&&-c_s^4
\Bigg(
      (\mathbf{k_1}\cdot\mathbf{k_2})(\mathbf{k_3}\cdot\mathbf{k_4})\bigg(\mathcal{F}_2(-k_{12},k_1,k_2,k_{12},k_3,k_4)-\mathcal{F}_2(-k_{12},-k_1,-k_2,k_{12},k_3,k_4)\bigg)
      \nonumber\\&&\quad\quad\quad\quad
      -2(\mathbf{k_1}\cdot\mathbf{k_2})(\mathbf{k_{34}}\cdot
\mathbf{k_4})\bigg(\mathcal{F}_2(-k_{12},k_1,k_2,k_3,k_4,k_{12})-\mathcal{F}_2(-k_{12},-k_1,-k_2,k_3,k_4,k_{12})\bigg)
      \nonumber\\&&\quad\quad\quad\quad
      -2(\mathbf{k_{12}}\cdot\mathbf{k_2})(\mathbf{k_3}\cdot\mathbf{k_4})\bigg(\mathcal{F}_2(k_1,k_2,-k_{12},k_{12},k_3,k_4)-\mathcal{F}_2(-k_1,-k_2,-k_{12},k_{12},k_3,k_4)\bigg)
      \nonumber\\&&\quad\quad\quad\quad
      +4(\mathbf{k_{12}}\cdot\mathbf{k_2})
(\mathbf{k_{34}}\cdot\mathbf{k_4})\bigg(\mathcal{F}_2(k_1,k_2,-k_{12},k_3,k_4,k_{12})-\mathcal{F}_2(-k_1,-k_2,-k_{12},k_3,k_4,k_{12})\bigg)
\Bigg)\Bigg]
\nonumber\\&&
\,\,
+23\,\mathrm{permutations\, of}\{k_1,k_2,k_3,k_4\},
\label{SETrispectrumPHI_ad}
\end{eqnarray}
where $\mathbf{K}=\mathbf{k}_1+\mathbf{k}_2+\mathbf{k}_3+\mathbf{k}_4$ and the information of the shape of the trispectrum
is encoded in the $\mathcal{F}_i$ functions defined by
Eqs.~(\ref{f1_express}), (\ref{f2_express}),
(\ref{f3_express}) and (\ref{f4_express}) in Appendix \ref{Fifc}.
In the previous expression, ``permutations" denote the other twenty three terms
that result from the permutations of $\{k_1,k_2,k_3,k_4\}$
in the preceding term.

The mixed component of the four-point function
coming from the scalar exchange diagrams is
\begin{eqnarray}
\langle\Omega|&&\!\!\!\!\!\!\!\!Q_\sigma(0,\mathbf{k_1})
Q_\sigma(0,\mathbf{k_2})Q_s(0,\mathbf{k_3})
Q_s(0,\mathbf{k_4})|\Omega\rangle^{SE} +5\,{\rm terms}
=(2\pi)^3\delta^{(3)}(\mathbf{K})
\frac{H^4}{16 X_0 c_s^9(k_1k_2k_3k_4)^\frac{3}{2}}\times
\nonumber\\
&&\Bigg[
-10 \bigg(\mathcal{F}_1(k_1,k_2,-k_{12},k_3,k_4,k_{12})-\mathcal{F}_1(-k_1,-k_2,-k_{12},k_3,k_4,k_{12})\bigg)
\nonumber\\&&
\,\,\,\,\,
+2 c_s^2
\Bigg(-
     (\mathbf{k_3}\cdot\mathbf{k_4})\bigg(\mathcal{F}_3(k_1,k_2,-k_{12},k_{12},k_3,k_4)-\mathcal{F}_3(-k_1,-k_2,-k_{12},k_{12},k_3,k_4)\bigg)
      \nonumber\\&&\quad\quad\quad\quad
      +4 (\mathbf{k_{34}}\cdot\mathbf{k_4})\bigg(\mathcal{F}_3(k_1,k_2,-k_{12},k_3,k_4,k_{12})-\mathcal{F}_3(-k_1,-k_2,-k_{12},k_3,k_4,k_{12})\bigg)
      \nonumber\\&&\quad\quad\quad\quad
      -(\mathbf{k_1}\cdot\mathbf{k_2})\bigg(\mathcal{F}_4(-k_{12},k_1,k_2,k_3,k_4,k_{12})-\mathcal{F}_4(-k_{12},-k_1,-k_2,k_3,k_4,k_{12})\bigg)
      \nonumber\\&&\quad\quad\quad\quad
      +4(\mathbf{k_{12}}\cdot\mathbf{k_2})\bigg(\mathcal{F}_4(k_1,k_2,-k_{12},k_3,k_4,k_{12})-\mathcal{F}_4(-k_1,-k_2,-k_{12},k_3,k_4,k_{12})\bigg)
\Bigg)
\nonumber\\
&&-2 c_s^4
\Bigg(
      (\mathbf{k_1}\cdot\mathbf{k_2})(\mathbf{k_3}\cdot\mathbf{k_4})\bigg(\mathcal{F}_2(-k_{12},k_1,k_2,k_{12},k_3,k_4)-\mathcal{F}_2(-k_{12},-k_1,-k_2,k_{12},k_3,k_4)\bigg)
      \nonumber\\&&\quad\quad\quad\quad
      +4(\mathbf{k_{12}}\cdot\mathbf{k_2})
(\mathbf{k_{34}}\cdot\mathbf{k_4})\bigg(\mathcal{F}_2(k_1,k_2,-k_{12},k_3,k_4,k_{12})-\mathcal{F}_2(-k_1,-k_2,-k_{12},k_3,k_4,k_{12})\bigg)
\Bigg)\Bigg]
\nonumber\\&&
\,\,
+23\,\mathrm{permutations\, of}\{k_1,k_2,k_3,k_4\}.
\label{SETrispectrumPHI_mixed}
\end{eqnarray}

Finally, the purely entropic four-point function
coming from the scalar exchange diagrams is
\begin{eqnarray}
\langle\Omega|&&\!\!\!\!\!\!\!\!Q_s(0,\mathbf{k_1})
Q_s(0,\mathbf{k_2})Q_s(0,\mathbf{k_3})
Q_s(0,\mathbf{k_4})|\Omega\rangle^{SE}
=(2\pi)^3\delta^{(3)}(\mathbf{K})
\frac{H^4}{16 X_0 c_s^9(k_1k_2k_3k_4)^\frac{3}{2}}\times
\nonumber\\
&&\Bigg[
-\bigg(\mathcal{F}_1(k_1,k_2,-k_{12},k_3,k_4,k_{12})-\mathcal{F}_1(-k_1,-k_2,-k_{12},k_3,k_4,k_{12})\bigg)
\nonumber\\&&
\,\,\,\,\,
+c_s^2
\Bigg(
     (\mathbf{k_3}\cdot\mathbf{k_4})\bigg(\mathcal{F}_3(k_1,k_2,-k_{12},k_{12},k_3,k_4)-\mathcal{F}_3(-k_1,-k_2,-k_{12},k_{12},k_3,k_4)\bigg)
      \nonumber\\&&\quad\quad\quad\quad
      +2(\mathbf{k_{34}}\cdot\mathbf{k_4})\bigg(\mathcal{F}_3(k_1,k_2,-k_{12},k_3,k_4,k_{12})-\mathcal{F}_3(-k_1,-k_2,-k_{12},k_3,k_4,k_{12})\bigg)
      \nonumber\\&&\quad\quad\quad\quad
      +(\mathbf{k_1}\cdot\mathbf{k_2})\bigg(\mathcal{F}_4(-k_{12},k_1,k_2,k_3,k_4,k_{12})-\mathcal{F}_4(-k_{12},-k_1,-k_2,k_3,k_4,k_{12})\bigg)
      \nonumber\\&&\quad\quad\quad\quad
      +2(\mathbf{k_{12}}\cdot\mathbf{k_2})\bigg(\mathcal{F}_4(k_1,k_2,-k_{12},k_3,k_4,k_{12})-\mathcal{F}_4(-k_1,-k_2,-k_{12},k_3,k_4,k_{12})\bigg)
\Bigg)
\nonumber\\
&&-c_s^4
\Bigg(
      (\mathbf{k_1}\cdot\mathbf{k_2})(\mathbf{k_3}\cdot\mathbf{k_4})\bigg(\mathcal{F}_2(-k_{12},k_1,k_2,k_{12},k_3,k_4)-\mathcal{F}_2(-k_{12},-k_1,-k_2,k_{12},k_3,k_4)\bigg)
      \nonumber\\&&\quad\quad\quad\quad
     + 2(\mathbf{k_1}\cdot\mathbf{k_2})(\mathbf{k_{34}}\cdot
\mathbf{k_4})\bigg(\mathcal{F}_2(-k_{12},k_1,k_2,k_3,k_4,k_{12})-\mathcal{F}_2(-k_{12},-k_1,-k_2,k_3,k_4,k_{12})\bigg)
      \nonumber\\&&\quad\quad\quad\quad
      +2(\mathbf{k_{12}}\cdot\mathbf{k_2})(\mathbf{k_3}\cdot\mathbf{k_4})\bigg(\mathcal{F}_2(k_1,k_2,-k_{12},k_{12},k_3,k_4)-\mathcal{F}_2(-k_1,-k_2,-k_{12},k_{12},k_3,k_4)\bigg)
      \nonumber\\&&\quad\quad\quad\quad
      +4(\mathbf{k_{12}}\cdot\mathbf{k_2})
(\mathbf{k_{34}}\cdot\mathbf{k_4})\bigg(\mathcal{F}_2(k_1,k_2,-k_{12},k_3,k_4,k_{12})-\mathcal{F}_2(-k_1,-k_2,-k_{12},k_3,k_4,k_{12})\bigg)
\Bigg)\Bigg]
\nonumber\\&&
\,\,
+23\,\mathrm{permutations\, of}\{k_1,k_2,k_3,k_4\}.
\label{SETrispectrumPHI_en}
\end{eqnarray}
The remaining possible four-point functions, $\langle\Omega|Q_\sigma Q_\sigma Q_\sigma
Q_s|\Omega\rangle^{SE}$ and $\langle\Omega|Q_\sigma Q_s Q_s
Q_s|\Omega\rangle^{SE}$, are zero at leading order in the slow-roll expansion and in the small sound speed limit.

As in the previous section,
the connected four-point function of the curvature perturbation
$\mathcal{R}$ is related with the four-point function of the field
perturbation as
\footnote{Where again we ignore terms that can be expressed by copies of 
the power spectrum by using Wick's theorem.}
\begin{eqnarray}
\langle\Omega|\mathcal{R} (\mathbf{k_1})
\mathcal{R} (\mathbf{k_2})
\mathcal{R} (\mathbf{k_3}) \mathcal{R} (\mathbf{k_4})
|\Omega\rangle^{SE}
&=& \mathcal{A}_{\sigma}^4
\langle\Omega|Q_\sigma (\mathbf{k_1})
Q_\sigma (\mathbf{k_2})
Q_\sigma (\mathbf{k_3}) Q_\sigma (\mathbf{k_4})
|\Omega\rangle^{SE}\nonumber\\
&&+\mathcal{A}_{\sigma}^2 \mathcal{A}_s^2
\left(\langle\Omega|Q_\sigma (\mathbf{k_1})
Q_\sigma (\mathbf{k_2})
Q_s (\mathbf{k_3}) Q_s (\mathbf{k_4})
|\Omega\rangle^{SE}
+5\,  {\rm perms}
\right)\nonumber\\
&&+\mathcal{A}_s^4
\langle\Omega|Q_s (\mathbf{k_1})
Q_s (\mathbf{k_2})
Q_s (\mathbf{k_3}) Q_s (\mathbf{k_4})
|\Omega\rangle^{SE}\,,
\label{4point_curv_scalar_exchange}
\end{eqnarray}
Again, as in the case of the contact interaction trispectra, the scalar exchange trispectra obey
\begin{eqnarray}
\langle
Q_\sigma(\mathbf{k_1})Q_\sigma(\mathbf{k_2})
Q_\sigma(\mathbf{k_3}) Q_\sigma(\mathbf{k_4})\rangle^{SE}
+
\langle
Q_s (\mathbf{k_1})Q_s (\mathbf{k_2})
Q_s (\mathbf{k_3}) Q_s (\mathbf{k_4})\rangle^{SE}
=
\langle
Q_\sigma(\mathbf{k_1})Q_\sigma(\mathbf{k_2})
Q_s (\mathbf{k_3}) Q_s (\mathbf{k_4})\rangle^{SE}
+ {\rm 5\,perm.}
\nonumber\\
\label{SE_simple_rel}
\end{eqnarray}
Using this fact Eq. (\ref{4point_curv_scalar_exchange}) can be simplified to
\begin{eqnarray}
\langle\mathcal{R}(\mathbf{k_1})\mathcal{R}(\mathbf{k_2})
\mathcal{R}(\mathbf{k_3}) \mathcal{R}(\mathbf{k_4})\rangle^{SE}
 &=& \mathcal{A}^4_\sigma\left(1+T_{\mathcal{RS}}^2\right)\bigg(
\langle
Q_\sigma(\mathbf{k_1})Q_\sigma(\mathbf{k_2})
Q_\sigma(\mathbf{k_3}) Q_\sigma(\mathbf{k_4})\rangle^{SE}
\nonumber\\
&&\qquad\qquad\qquad\quad
+T_{\mathcal{RS}}^2
\langle
Q_s (\mathbf{k_1})Q_s (\mathbf{k_2})
Q_s (\mathbf{k_3}) Q_s (\mathbf{k_4})\rangle^{SE}\bigg)\,.\label{4point_curv_scalar_exchangesimplified}
\end{eqnarray}

Finally, the total four-point function of $\mathcal{R}$, 
coming from the quantum four-point functions of the fields,
is the sum of the contact interaction and 
scalar exchange contributions as
\begin{eqnarray}
&&\!\!\!\!\!\!\!\!\!\!\!\!\!\!\!\!\!\!\!\!\!\!\!\!\!\!\langle\mathcal{R} (\mathbf{k_1})
\mathcal{R} (\mathbf{k_2})
\mathcal{R} (\mathbf{k_3}) \mathcal{R} (\mathbf{k_4})
\rangle^{total}
=
\langle\mathcal{R} (\mathbf{k_1})
\mathcal{R} (\mathbf{k_2})
\mathcal{R} (\mathbf{k_3}) \mathcal{R} (\mathbf{k_4})
\rangle^{CI}
+\langle\mathcal{R} (\mathbf{k_1})
\mathcal{R} (\mathbf{k_2})
\mathcal{R} (\mathbf{k_3}) \mathcal{R} (\mathbf{k_4})
\rangle^{SE}\nonumber\\
&&=
\mathcal{A}^4_\sigma\left(1+T_{\mathcal{RS}}^2\right)
\bigg[
\langle
Q_\sigma(\mathbf{k_1})Q_\sigma(\mathbf{k_2})
Q_\sigma(\mathbf{k_3}) Q_\sigma(\mathbf{k_4})\rangle^{CI}
+
\langle
Q_\sigma(\mathbf{k_1})Q_\sigma(\mathbf{k_2})
Q_\sigma(\mathbf{k_3}) Q_\sigma(\mathbf{k_4})\rangle^{SE}
\nonumber\\
&&\qquad\qquad\qquad\qquad
+T_{\mathcal{RS}}^2
\bigg(
\langle
Q_s (\mathbf{k_1})Q_s (\mathbf{k_2})
Q_s (\mathbf{k_3}) Q_s (\mathbf{k_4})\rangle^{CI}
+
\langle
Q_s (\mathbf{k_1})Q_s (\mathbf{k_2})
Q_s (\mathbf{k_3}) Q_s (\mathbf{k_4})\rangle^{SE}\bigg)
\bigg]
\,,
\label{4point_curv_total}
\end{eqnarray}
where the different terms can be found in Eqs. (\ref{4point_pureadiabatic}), (\ref{SETrispectrumPHI_ad}), (\ref{4point_pureentropy}) and (\ref{SETrispectrumPHI_en}). This constitutes one of the main results of this work.

From Eq. (\ref{4point_curv_total}) we can read
the exact momentum dependence of the trispectrum
which is related with the four-point function
as
\begin{equation}
\langle\mathcal{R}(\mathbf{k}_1)
\mathcal{R}(\mathbf{k}_2)\mathcal{R}(\mathbf{k}_3)
\mathcal{R}(\mathbf{k}_4)\rangle^{total}
=(2\pi)^3\delta^{(3)}(\mathbf{K})
T_\mathcal{R}(\mathbf{k}_1,\mathbf{k}_2,\mathbf{k}_3,
\mathbf{k}_4).
\end{equation}

In the next section we shall study the shape
of the trispectrum in more detail.

\section{\label{sec:shape}The shape of the trispectrum
and the non-linearity parameter $\tau_{NL}$}

In this section, we will study the shape dependence of the different trispectra calculated in the previous sections. We will consider the so-called equilateral configuration.
In the next subsection, we shall discuss some consistency relations that
apply for the multi-field DBI trispectrum. After that, we will compute
the non-linearity parameter $\tau_{NL}$ and we shall plot the different
contributions (pure adiabatic and pure entropy) to the total $\tau_{NL}$
both in the equilateral case and in a non-equilateral configuration. It
is worth noting that from Eqs.~(\ref{CI_simple_rel})
and (\ref{SE_simple_rel}), the contribution from the mixed
component can be given from the other two contributions.

\subsection{Consistency relations}

Here, we discuss whether the consistency relations
obtained in the single field DBI inflation model can hold also
for the multi-field DBI inflation model. In the following, as in  \cite{Arroja:2009pd},
we will define the non-linearity parameter, which
naively parameterizes the size of the trispectrum as
\begin{eqnarray}
\tau_{NL}(\mathbf{k}_1,\mathbf{k}_2,\mathbf{k}_3,\mathbf{k}_4)&=&\left(\frac{4\epsilon c_s}{H^2(1+T_{\mathcal{RS}}^2)}\right)^3
T_\mathcal{R}(\mathbf{k}_1,\mathbf{k}_2,\mathbf{k}_3,\mathbf{k}_4) \Pi_{i=1}^4k_i^3\nonumber\\&&\!\!\!\!\!\times
\bigg[
      \left(k_1^3k_2^3+k_3^3k_4^3\right)\left(k_{13}^{-3}+k_{14}^{-3}\right)
      +\left(k_1^3k_4^3+k_2^3k_3^3\right)\left(k_{12}^{-3}+k_{13}^{-3}\right)
      +\left(k_1^3k_3^3+k_2^3k_4^3\right)\left(k_{12}^{-3}+k_{14}^{-3}\right)
\bigg]^{-1},\nonumber\\
\label{taunl_def}
\end{eqnarray}
and we discuss behaviours of $\tau_{NL}$.

First, we discuss the consistency relation
in the squeezed limit. As first pointed out
by Maldacena \cite{Maldacena:2002vr} and
Seery \emph{et al.} \cite{Seery:2006vu},
all higher order correlators of $\mathcal{R}$
in single field inflation obey consistency relations
when one of the momentum vector is very small.
Let us consider the limit $\mathbf{k}_1\rightarrow0$,
which implies that the corresponding mode
$\mathcal{R}(\mathbf{k}_1)$ leaves the horizon
well before all other modes leave the horizon.
By the time the remaining modes exit the horizon,
$\mathcal{R}(\mathbf{k}_1)$
will be frozen as a super-horizon mode
and its only effect is to deform the background if
the mode with $\mathbf{k}_1$ which leaves the horizon
earlier than the others is the adiabatic mode ($\sigma$).
Then in this case, the trispectrum can be expressed as \cite{Seery:2006vu}
\begin{equation}
\langle\mathcal{R}(\mathbf{k}_1)\mathcal{R}(\mathbf{k}_2)
\mathcal{R}(\mathbf{k}_3)\mathcal{R}(\mathbf{k}_4)\rangle
\rightarrow-\tilde {\cal P}_\mathcal{R}(k_1)\frac{d}{Hdt}
\langle\mathcal{R}(\mathbf{k}_2)\mathcal{R}(\mathbf{k}_3)
\mathcal{R}(\mathbf{k}_4)\rangle,
\label{consistencyrel2}
\end{equation}
where the power spectrum
$\tilde {\cal P}_\mathcal{R}(k)$ is given by
$\langle\mathcal{R}(\mathbf{k}_1)\mathcal{R}(\mathbf{k}_2)
\rangle=(2\pi)^3\delta^{(3)}(\mathbf{k}_1+\mathbf{k}_2)
\tilde {\cal P}_\mathcal{R}(k_1)$.
Using the fact that the bispectrum is $B_\zeta \propto c_s^{-2}{\cal \tilde P}_{\cal R}^2$, it follows that in the squeezed limit the trispectrum is ${\cal T}_R \propto (\epsilon c_s^{-2}{\cal \tilde P}_{\cal R}^3)$. Using $\tilde{P}_{\cal R} \propto (c_s \epsilon)^{-1}$, one finds that the total non-linearity parameter $\tau_{NL}$ should be at most of order $\epsilon c_s^{-2}$ (i.e. under our approximation it should vanish) when any of the momentum vectors goes to zero. From Eqs.~(\ref{4point_pureadiabatic}),
(\ref{4point_pureentropy}), (\ref{4point_mixed}),
(\ref{SETrispectrumPHI_ad}), (\ref{SETrispectrumPHI_mixed})
and (\ref{SETrispectrumPHI_en}), we can show that $\tau_{NL}$ scales as $\mathcal{O}(k_1^{2})$, which means the consistency relation in the squeezed limit
holds for the trispectrum in multi-field DBI inflation models.

However, if the mode with $\mathbf{k}_1$ is the entropy mode ($s$), this
explanation can not be applied. Instead, the fact that at leading order
in slow-roll, the third-order
action is written as Eq.~(\ref{two_field_lead_dbi_thir})
and there are no three point functions like ($sss$)
and ($s \sigma \sigma$) turns out to be crucial.
Actually, if one of the entropy modes leaves the horizon
and becomes classical, the combination of the remaining
three variables are ($sss$) for pure entropy component
and ($s \sigma \sigma$) for mixed component, respectively.
As mentioned above, since these three-point correlations
are absent at leading order in slow roll, the trispectrum
vanishes in this limit.

Similarly, we can discuss the consistency relation
in the counter-collinear limit for the trispectra
coming from the scalar exchange diagram.
According to \cite{Seery:2008ax},
in the limit where the momentum of the scalar mode
that is exchanged goes to zero,
one can find a simple relation
between the scalar exchange trispectrum and the power spectrum,
in a similar way to Maldacena's consistency relations
\cite{Maldacena:2002vr} (see also
\cite{Creminelli:2004yq,Huang:2006eh,Cheung:2007sv})
since one can treat this mode as a background
again if this mode is an adiabatic mode.
Let us suppose that the momentum
of the exchanged adiabatic particle is $k_{12}$ and
that $k_{12}\ll k_1\approx k_2,k_3\approx k_4$.
Then the mode associated with this scalar particle
will cross the horizon much before the other $k_i$ modes,
where $i=1,\ldots,4$, and it only rescale
the spatial background where the $k_i$ modes exist.
Then, in the limit $\mathbf{k}_{12}\rightarrow0$,
the following relation should hold if the exchange mode is an adiabatic mode:
\begin{equation}
\langle\mathcal{R}(\mathbf{k}_1)\mathcal{R}(\mathbf{k}_2)
\mathcal{R}(\mathbf{k}_3)\mathcal{R}(\mathbf{k}_4)\rangle ^{SE}
\rightarrow (2\pi)^3\delta^{(3)}(\mathbf{K})(n_\mathcal{R}-1)^2
\tilde {\cal P}_\mathcal{R}(k_{12})
\tilde {\cal P}_\mathcal{R}(k_{1})
\tilde {\cal P}_\mathcal{R}(k_{3}).
\label{consistencyrel}
\end{equation}
The previous equation implies that in the counter-collinear limit, the scalar exchange non-linearity parameter $\tau_{NL}^{SE}$ is of order $\epsilon^2$ (at leading order in our approximations this is equivalent to say that $\tau_{NL}^{SE}$ should vanish).
From Eqs.~(\ref{SETrispectrumPHI_ad}),
(\ref{SETrispectrumPHI_mixed})
and (\ref{SETrispectrumPHI_en}), the non-linearity parameter $\tau_{NL}$
is shown to scale as $\mathcal{O}(k_{12}^4)$.
Therefore,  the consistency relation in
the counter-collinear limit holds for the trispectrum coming from
the scalar exchange diagram generated from
the multi-field DBI inflation model.
For the trispectrum coming from the other diagram,
since there is no $k_{12}$ dependence,
this consistency relation holds trivially.

For the pure adiabatic, pure entropy and part of the mixed component,
the exchanged mode is always an adiabatic mode. Thus we can apply the
above argument to show that $\tau_{NL}$ vanishes in this limit. On the other hand,
if the mode with $\mathbf{k}_{12}$ is the entropy mode, which happens for
the mixed component, we cannot apply the above arguments. However, in this case,
the four-point function becomes
a square of the two-point function of ($s \sigma$).
As was shown by Eq.~(\ref{two_field_lead_dbi_sec}),
there is no two-point correlation between the adiabatic
and entropy mode at leading order in slow-roll.
This explains the consistency relation
in the counter-collinear limit for the remaining part
of the mixed component.

Now that we have checked the two important consistency
relations, in the next subsection, we will concentrate on
a concrete and best studied configuration
in single field DBI inflation model, the
so-called equilateral configuration
\cite{Seery:2006vu,Huang:2006eh,Arroja:2009pd,Chen:2009bc}.

\subsection{\label{subsec:Shape}The shape of the trispectrum
and $\tau_{NL}$ in the equilateral configuration}

In this subsection, we shall study the shape of the trispectrum
in the limit where all the four momentum vectors
have the same magnitude $k$.
If we denote the angle between $\mathbf{k}_i$ and
$\mathbf{k}_j$ by
$\theta_{ij}$, then in this configuration, we have
$\cos(\theta_{12})=\cos(\theta_{34})\equiv\cos(\theta_3)$,
$\cos(\theta_{23})=\cos(\theta_{14})\equiv\cos(\theta_1)$,
$\cos(\theta_{13})=\cos(\theta_{24})\equiv\cos(\theta_2)$ and
$\cos(\theta_1)+\cos(\theta_2)+\cos(\theta_3)=-1$
due to momentum
conservation. Contrary to
what happens in the case of the bispectrum,
the equilateral configuration conditions
do not fix all degrees of freedom (dof)
required to describe the shape of the trispectrum
and we are left with two angular dof \cite{Seery:2006vu}.

\begin{figure}[t]
\scalebox{1
}{
\centerline{
\includegraphics{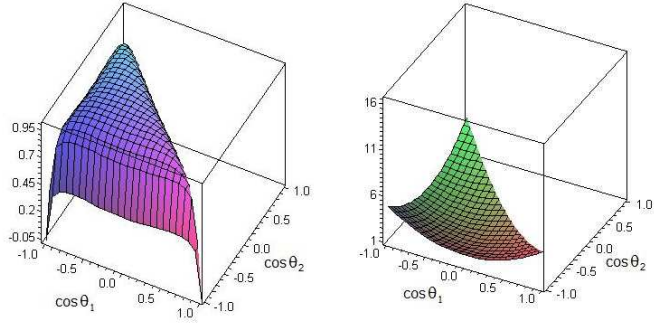}}}
\caption{The shape of the scalar exchange adiabatic (l.h.s panel)
and pure entropy (r.h.s. panel)
trispectra as functions of the variables
$\cos \theta_1$ and $\cos \theta_2$. The amplitude of the trispectra at the point $\cos\theta_1=\cos\theta_2=-1/3$ has been normalized to unity.}
\label{SEtrispecta}
\end{figure}

In the equilateral configuration,
the contact interaction trispectrum is constant and
only the scalar exchange trispectrum depends on the remaining angular dof, that we choose to be $\theta_1$ and $\theta_2$.
The plots of the shape of the scalar exchange trispectrum
coming from the purely adiabatic component (the first term in
Eq.~(\ref{4point_curv_scalar_exchange}))
and
the purely entropy component
(the last term in Eq.~(\ref{4point_curv_scalar_exchange}))
can be found in Fig. \ref{SEtrispecta}. Using Eq. (\ref{4point_curv_scalar_exchangesimplified}), one can easily see that,
for $T_{\mathcal{R}\mathcal{S}} \ll 1$
the purely adiabatic component dominates while
for $T_{\mathcal{R}\mathcal{S}} \gg 1$,
the purely entropy component dominates.
Because the momentum dependence of the trispectrum
changes depending on the value of
$T_{\mathcal{R}\mathcal{S}}$, the trispectrum
can distinguish multi-field DBI inflation
models from single field DBI inflation models
if $T_{\mathcal{R}\mathcal{S}}$ is significantly large.
The exact analytical expressions for the different contributions in the scalar exchange trispectrum can be found
in Appendix \ref{TDBI}.

\begin{figure}[t]
\scalebox{0.9}{
\centerline{
\includegraphics{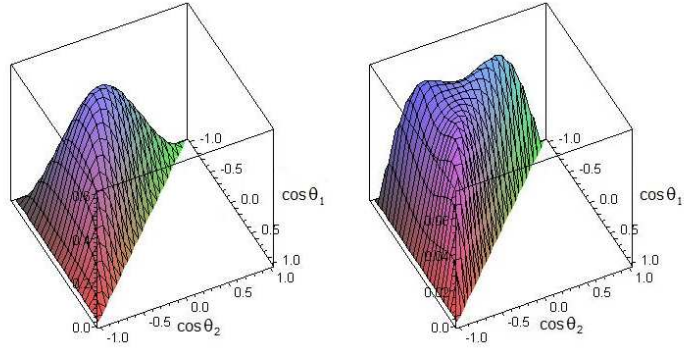}}}
\caption{Plots of the non-linearity parameter $\tau_{NL}^{SE}$
 calculated from the scalar exchange adiabatic (l.h.s. panel)
and pure entropy (r.h.s.panel) trispectra
as functions of the variables
$\cos \theta_1$ and $\cos \theta_2$.
The amplitude of the first plot was rescaled by $c_s^4(1+T_{\mathcal{RS}}^2)^3$, while the amplitude of the r.h.s. plot was rescaled by $c_s^4(1+T_{\mathcal{RS}}^2)^3/T_{\mathcal{RS}}^4$.}
\label{tauNLSE}
\end{figure}

Even though the trispectrum includes all the information, for practical purposes it is more convenient to use another quantity $\tau_{NL}$  defined in Eq.~(\ref{taunl_def}).
In Fig. \ref{tauNLSE}, we plot the non-linearity parameter
$\tau_{NL}^{SE}$ for the purely adiabatic
and
purely entropy components
calculated from the respective terms in the scalar exchange trispectrum, Eq. (\ref{4point_curv_scalar_exchange}).

\begin{figure}[t]
\scalebox{0.8}{
\centerline{
\includegraphics{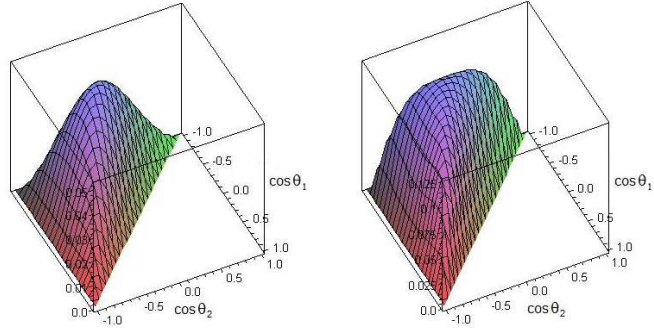}}}
\caption{L.h.s. panel: Plot of the non-linearity parameter $\tau_{NL}^{CI}$ calculated from the contact interaction pure entropy trispectrum as a function of $\cos\theta_1$ and $\cos\theta_2$. R.h.s. panel: Plot of the sum of $\tau_{NL}^{CI}$ and $\tau_{NL}^{SE}$ coming from the pure entropy trispectrum as a function of $\cos\theta_1$ and $\cos\theta_2$. The amplitude of the plots was rescaled by $c_s^4(1+T_{\mathcal{RS}}^2)^3/T_{\mathcal{RS}}^4$.
}
\label{tauNLCIplot}
\end{figure}

In Fig. \ref{tauNLCIplot}, we plot the non-linearity parameter
$\tau_{NL}^{CI}$ for the purely entropy component of the contact interaction trispectrum. The shape of the surfaces for the purely adiabatic and mixed components are the same as the one for the purely entropy component. However, the amplitudes are different, in particular, for the purely adiabatic plot, $\tau_{NL}^{CI}$ is negative. On the r.h.s. of Fig. \ref{tauNLCIplot}, we plot the sum of the non-linearity parameters $\tau_{NL}^{CI}$ and $\tau_{NL}^{SE}$ that come from purely entropic perturbations. If the transfer coefficient is large this will be the dominant shape in the total non-linearity parameter.

Notice that since the factor $k_{ij}^3$ is multiplying the trispectrum,
in terms of $\tau_{NL}$,
we confirm the consistency relation in
the counter-collinear limit of the trispectrum
coming from pure adiabatic, mixed and pure entropy components.
When $\cos\theta_1=-1$ or $\cos\theta_2=-1$ or
$\cos\theta_3=-1$ (on the diagonal lines in the plots of
Fig. \ref{tauNLSE}, when $\cos\theta_2=-\cos\theta_1$)
we can see the non-linearity parameters are zero.

In order to quote a number for the value of the non-linearity parameter, we need to
fix the remaining two degrees of freedom ($\theta_1$ and $\theta_2$)
and in this way fully specify the momentum vectors' configuration.
In the single field DBI model, there are good reasons to believe that the maximum of $\tau_{NL}$ occurs for the equilateral
configuration with
$\cos(\theta_1)=\cos(\theta_2)=\cos(\theta_3)=-1/3$
\cite{Arroja:2009pd} \footnote{However, the definite proof that this is the case is still missing}.
In the multi-field case, for some contributions (like for example, $\tau_{NL}^{SE}$ coming from the pure entropy trispectrum, as seen on the r.h.s. panel of Fig. \ref{tauNLSE}) the non-linearity parameter is not maximized for this particular angular configuration. However, for the sake of comparison with the single field model, we shall obtain the non-linearity parameter in that configuration.

So, for the equilateral configuration, with $\cos(\theta_1)=\cos(\theta_2)=\cos(\theta_3)=-1/3$, we obtain
\begin{equation}
{\tau_{NL}}^{CI}_{4\sigma}\sim
-\frac{0.036}{c_s^4 (1+T_{\mathcal{R}\mathcal{S} }^2)^3},
\quad
{\tau_{NL}}^{CI}_{\sigma \sigma ss}\sim
\frac{0.016T_{\mathcal{R}\mathcal{S}}^2}
{c_s^4 (1+T_{\mathcal{R}\mathcal{S}}^2)^3},
 \quad
{\tau_{NL}}^{CI}_{4s}\sim
\frac{0.052 T_{\mathcal{R}\mathcal{S}}^4}
{c_s^4 (1+T_{\mathcal{R}\mathcal{S} }^2)^3}, \label{tauNLCI}
\end{equation}
where the different values of $\tau_{NL}$ correspond to the
three terms in the contact interaction trispectrum
coming from the purely adiabatic, mixed and purely entropic components in
Eq.~(\ref{4point_curv_cont}), respectively.

Similarly, from the scalar exchange trispectrum we obtain
\begin{equation}
{\tau_{NL}}^{SE}_{4\sigma} \sim
\frac{0.60}{c_s^4 (1+T_{\mathcal{R}\mathcal{S} }^2)^3},
\quad
{\tau_{NL}}^{SE}_{\sigma \sigma ss}\sim
\frac{0.67 T_{\mathcal{R}\mathcal{S}}^2}
{c_s^4 (1+T_{\mathcal{R}\mathcal{S} }^2)^3}
, \quad
{\tau_{NL}}^{SE}_{4s}\sim
\frac{0.071 T_{\mathcal{R}\mathcal{S}}^4}
{c_s^4 (1+T_{\mathcal{R}\mathcal{S} }^2)^3},
\label{tauNLSEeq}
\end{equation}
where again
the different values of $\tau_{NL}$ correspond to the
three terms in the scalar exchange interaction trispectrum
coming from the purely adiabatic, mixed and purely entropic components in
Eq.~(\ref{4point_curv_scalar_exchange}),
respectively.

It is worth mentioning that from
Eqs.~(\ref{CI_simple_rel})
and (\ref{SE_simple_rel}),
both for the contact interaction (Eq.~(\ref{tauNLCI}))
and the scalar exchange interaction (Eq.~(\ref{tauNLSEeq})),
we can obtain the relation
\begin{equation}
{\tau_{NL}}_{\sigma \sigma s s}= T_{\mathcal{R}\mathcal{S} }^2
{\tau_{NL}}_{4 \sigma}+\frac{1}{ T_{\mathcal{R}\mathcal{S} }^2}
{\tau_{NL}}_{4 s}\,.
\end{equation}

Therefore, we conclude that the total non-linearity
parameter $\tau_{NL}$ for the multi-field DBI inflation model
in the equilateral configuration with equal
angles between the momentum vectors is
\begin{equation}
{\tau_{NL}}^{tot}_{equi} \sim
\frac{0.56}{c_s^4 (1+T_{\mathcal{R}\mathcal{S} }^2)^3}+
\frac{0.68 T_{\mathcal{R}\mathcal{S}}^2}
{c_s^4 (1+T_{\mathcal{R}\mathcal{S} }^2)^3}+
\frac{0.12 T_{\mathcal{R}\mathcal{S}}^4}
{c_s^4 (1+T_{\mathcal{R}\mathcal{S} }^2)^3}\/.
\label{tauNLtot}
\end{equation}
It is worth noting that when there is no conversion of entropy perturbation into curvature perturbation, i.e. $T_{\mathcal{R}\mathcal{S}}=0$,
Eq.~(\ref{tauNLtot}) reproduces the result
obtained in the single field DBI inflation model
\cite{Arroja:2009pd}.
On the other hand,  when the transfer coefficient is large, i.e. $T_{\mathcal{R}\mathcal{S}}\gg1$, the final curvature perturbation is mainly of entropy origin and the last term in Eq.~(\ref{tauNLtot}) dominates
over the others.
\subsection{\label{subsec:tauNLnonequi}
The non-linearity parameter $\tau_{NL}$ in
a non-equilateral configuration}

As we have mentioned before, the equilateral configuration
is well investigated in the DBI inflation model.
However, since there are five degrees of freedom
to parameterize the shape of the trispectrum and
the expression for the trispectrum
is fairly complicated, non-equilateral configurations have attracted less attention. Because it has not been shown that the equilateral configuration maximizes the value of the non-linearity parameter, in this subsection we will consider a non-equilateral configuration.

In our previous work \cite{Arroja:2009pd},
we have studied the deviation from the equilateral configuration
by fixing a particular set of degrees of freedom
in the single field DBI inflation model case. Here, we apply the same
approach and we study the behaviour of $\tau_{NL}$
in a non-equilateral configuration.

\begin{figure}[t]
\scalebox{0.9}{
\centerline{
\includegraphics{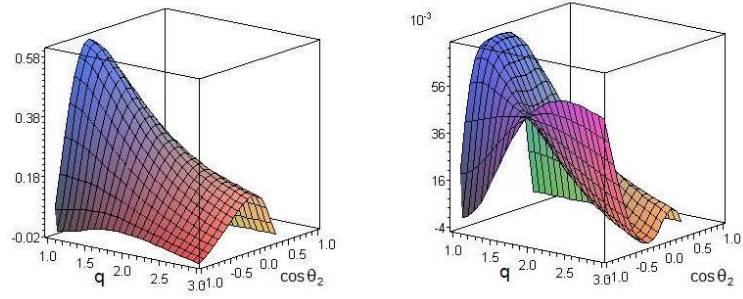}}}
\caption{
Plots of the non-linearity parameter $\tau_{NL}^{SE}$
calculated from the scalar exchange adiabatic (l.h.s. panel)
and pure entropy (r.h.s. panel)
trispectra for non-equilateral configurations
as functions of the momentum amplitude $q$ and $\cos \theta_2$.
The amplitude of the first plot was rescaled by
$c_s^4(1+T_{\mathcal{RS}}^2)^3$, while the amplitude of
the r.h.s. plot was rescaled by
$c_s^4(1+T_{\mathcal{RS}}^2)^3/T_{\mathcal{RS}}^4$.}
\label{tauNLSENonEqui}
\end{figure}

\begin{figure}[t]
\scalebox{0.9}{
\centerline{
\includegraphics{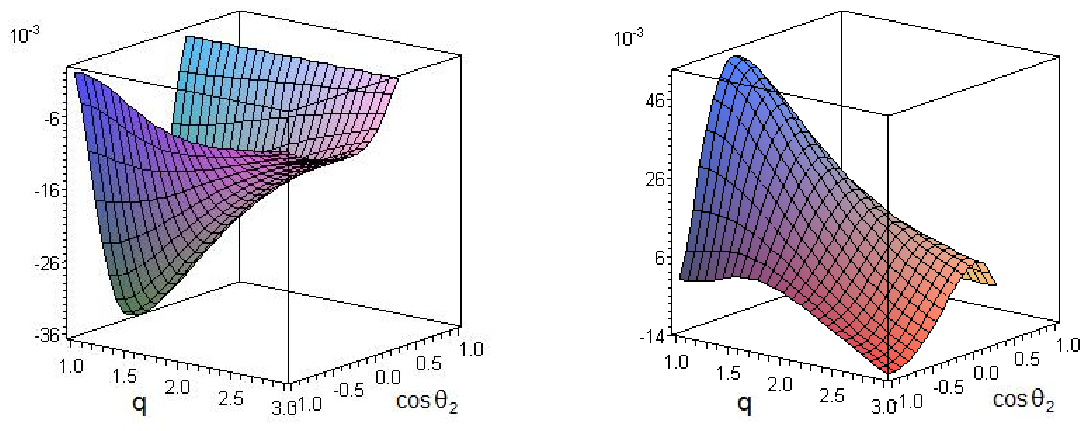}}}
\caption{
Plots of the non-linearity parameter $\tau_{NL}^{CI}$
calculated from the contact interaction adiabatic (l.h.s. panel) and pure entropy (r.h.s. panel)
trispectra for non-equilateral configurations
as functions of the momentum amplitude $q$ and $\cos \theta_2$.
The amplitude of the first plot was rescaled by
$c_s^4(1+T_{\mathcal{RS}}^2)^3$, while the amplitude of
the r.h.s. plot was rescaled by
$c_s^4(1+T_{\mathcal{RS}}^2)^3/T_{\mathcal{RS}}^4$.}
\label{tauNLCINonEqui}
\end{figure}

\begin{figure}[t]
\scalebox{0.8}{
\centerline{
\includegraphics{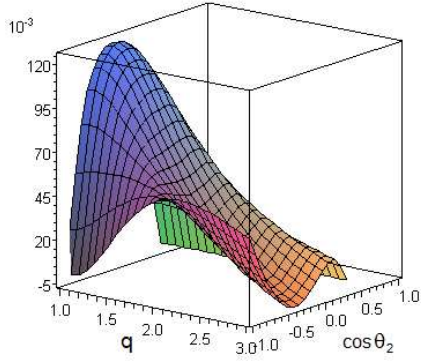}}}
\caption{
Plot of the sum of the non-linearity parameters $\tau_{NL}^{SE}$ and $\tau_{NL}^{CI}$
calculated from the pure entropy trispectra, for non-equilateral configurations,
as functions of the momentum amplitude $q$ and $\cos \theta_2$.
The amplitude of the plot was rescaled by
$c_s^4(1+T_{\mathcal{RS}}^2)^3/T_{\mathcal{RS}}^4$.}
\label{tauNLCISENonEqui}
\end{figure}

We consider configurations where
the two momentum vectors $\mathbf{k}_3$ and $\mathbf{k}_4$
have a different magnitude $q$
from the magnitude of $\mathbf{k}_1$ and
$\mathbf{k}_2$ that we normalized to unity.
 Then in this configuration, we have
$\cos(\theta_{34}) \equiv \cos(\theta_3),
\cos(\theta_{23})=\cos(\theta_{14}) \equiv \cos(\theta_1),
\cos(\theta_{13})=\cos(\theta_{24}) \equiv \cos(\theta_2),
\cos(\theta_{12})= q^2 (1+\cos(\theta_3))-1$.
Momentum conservation implies $\cos\theta_3=-(1/q)(\cos\theta_1+\cos\theta_2)-1$.
We will allow one of the angular dof to vary freely,
the other angular dof $\theta_1$ is fixed
by $\cos\theta_1=-1/3$ to the value that gives us the maximum
for the single field model.
In Fig. \ref{tauNLSENonEqui},
we plot the non-linearity parameter $\tau_{NL}^{SE}$
calculated from the scalar exchange adiabatic (l.h.s. panel)
and entropic (r.h.s. panel) trispectra,
respectively. The maximum appears at $q=1$ and $\cos(\theta_2)=-1/3$, which
supports the importance of the equilateral configuration.
In Fig. \ref{tauNLCINonEqui},
we plot the non-linearity parameter $\tau_{NL}^{CI}$
calculated from the contact interaction
adiabatic (l.h.s. panel) and entropic
(r.h.s. panel) trispectra,
respectively.
Comparing Figs. \ref{tauNLSENonEqui} and \ref{tauNLCINonEqui}, one can
see that around the maximum the purely adiabatic
contribution coming from the contact interaction is smaller than the corresponding one coming from the scalar exchange interaction. Only for the entropy component, the amplitudes of the scalar exchange and contact interaction trispectra are comparable. We plot the sum of these two terms in Fig. \ref{tauNLCISENonEqui}. If the transfer coefficient is large this will be the dominant shape.

\section{\label{sec:OBSERVATIONS}
Observational constraints}

Since we have obtained the full quantum trispectra
for each component for the equilateral configuration
with $\cos(\theta_1)=\cos(\theta_2)=\cos(\theta_3)=-1/3$,
in terms of $T_{\mathcal{R}\mathcal{S}}$,
we are in a position to discuss the relation between
$T_{\mathcal{R}\mathcal{S}}$ and observable quantities. It was shown in \cite{Langlois:2009ej} that
in the small sound speed limit ($c_s^2 \ll 1$),
the spectral index
of the scalar perturbations $n_{\mathcal{R}}$
in two-field DBI inflation model can be written as
\begin{eqnarray}
&&1-n_{\mathcal{R}} \simeq
\frac{\sqrt{3.1 |f_{NL}^{equil}| }r}
{4 \cos^3 \Theta}-\frac{\dot{f}}{H f} +
\alpha_* \sin (2 \Theta) + 2 \beta_* \sin^2 \Theta,
\label{ns_obs}\\
&&{\rm with}\quad
\alpha = \frac{\xi}{a H},\quad \beta= \frac{\chi}{2}
-\frac{\iota}{2}-\frac{1}{3 H^2}
\left( \mu_s^2 + \frac{\xi^2}{a^2}\right),
\quad
\Theta = \arcsin \left(\frac{T_{\mathcal{R} \mathcal{S}}}
{\sqrt{1+T_{\mathcal{R} \mathcal{S}}^2}}\right),
\end{eqnarray}
where $\iota$ and $\chi$ are the slow-roll parameters
defined in Eq.~(\ref{slow_roll_parameters}).
$\xi$ is the coupling between the adiabatic
and entropy perturbation and $\mu_s$ is the mass
of the entropy perturbation, which are model dependent
quantities.
In order to obtain
Eq.~(\ref{ns_obs}), they used the following
relations:
\begin{eqnarray}
&&r=\frac{\mathcal{P}_T}{\mathcal{P}_\mathcal{R}}
=16 \epsilon c_s \cos^2 \Theta,\label{r_obs}\\
&&f_{NL} ^{equi} \simeq -\frac{1}{3.1 c_s^2}
\cos^2 \Theta,\label{fnl_obs}
\end{eqnarray}
where $r$ is the tensor to scalar ratio and $f_{NL}^{equi}$ parameterizes the size of the bispectrum in the equilateral configuration.

In the ultra-violet (UV) DBI model ($\dot{f} > 0$) and without entropy
transfer, i.e. $T_{\mathcal{R} \mathcal{S}} =0$
(or equivalently $\sin \Theta =0$),
Eq.~(\ref{ns_obs}) gives the well known stringent
constraint as
\begin{eqnarray}
r > \frac{4}{\sqrt{3.1 |f_{NL} ^{equi}|}}(1-n_{\mathcal{R}}),
\label{single_obs_cost}
\end{eqnarray}
by which one obtains $r \geq 10^{-3}$
if one uses the observational constraints on $f_{NL} ^{equi}$
and $1-n_{\mathcal{R}}$ from WMAP5 data \cite{Komatsu:2008hk}.
On the other hand, since for general compact internal
spaces, suggested by string theory, there is
a geometrical constraint that limits $r$ to be
$r < 10^{-7}$ \cite{Lidsey:2007gq}, the single field UV DBI inflation
model is already excluded by the observational data.

However, if we take into account the transfer
from the entropy perturbation, the last two terms
in Eq.~(\ref{ns_obs}) are no longer negligible.
Furthermore, from Eqs.~(\ref{r_obs}) and (\ref{fnl_obs}),
both $r$ and $f_{NL} ^{equi}$ are suppressed
by $\cos ^2 \Theta$ which scales as
$ 1/T_{\mathcal{R}\mathcal{S}}^2$ for
$T_{\mathcal{R}\mathcal{S}} \gg 1$.
Because of these factors,
the constraint given by Eq.~(\ref{single_obs_cost})
cannot be applied and it is possible to satisfy the
geometrical constraint $r < 10^{-7}$
if $T_{\mathcal{R}\mathcal{S}} \gg 1$ \cite{Langlois:2009ej}.
In such a case, since the last term of Eq. (\ref{tauNLtot})
will dominate over the other terms,
$\tau_{NL} ^{equi}$ can be expressed as
$\tau_{NL} ^{equi} \sim 1.14 (f_{NL} ^{equi})^2
T_{\mathcal{R} \mathcal{S}}^2$.
Thus for a given $f_{NL}^{equi}$, $\tau_{NL}^{equi}$ is enhanced for a large
transfer $T_{\mathcal{RS}} \gg 1$, while the global magnitude
of $\tau_{NL}$ is reduced, going like $1/T_{\mathcal{RS}}^2$
in this case.

\section{\label{sec:CONCLUSIONS}
Conclusions}

In this work, we have obtained the full quantum trispectrum
of the primordial curvature perturbation at leading order
in the slow-roll expansion and in the small sound speed limit
in multi-field DBI inflation models. Previously, we had obtained the contact interaction trispectrum \cite{Mizuno:2009cv}. In this work we completed the calculation of
the quantum
trispectrum including the scalar exchange trispectrum.

We studied the momentum dependence of the trispectrum
by separating it into the purely adiabatic, mixed and purely
entropic components. For general configurations, as in the single field
model, we confirmed that the consistency relations in the squeezed
and in the counter-collinear limit hold. If a mode
that crosses the horizon much before the other modes is an adiabatic one,
these can be explained by the usual argument that the mode only changes the background
(just like in the single field models). However, if the mode that crosses the horizon first is an entropic one,
the consistency relations hold because of the absence of the two-point and
three-point correlations among the modes remaining under
the horizon scale. Essentially, the absence of these correlations is attributed to
the particular properties of the DBI action discussed in
\cite{Mizuno:2009cv}.

In the equilateral configuration,
we examined the momentum dependence of the scalar exchange
trispectra as functions of the two remaining
angular degrees of freedom. We found that the shape of the
adiabatic, mixed, and purely entropic trispectra
are different. This means that if the amount
of the transfer from the entropy perturbation to the final curvature perturbation is significantly
large, the trispectrum can distinguish multi-field DBI
inflation models from single field DBI inflation models.
In the equilateral configuration with $\cos\theta_1=\cos\theta_2=\cos\theta_3=-1/3$, we showed that for the contributions for $\tau_{NL}$ coming from the adiabatic and mixed trispectra, the scalar exchange diagram is more important than the contact interaction diagram, while for the pure entropy contribution both diagrams are equally important. This suggests that for quantitative discussions, the analysis done in Ref.~\cite{Gao:2009gd}
taking into account only the scalar exchange diagram
is insufficient, at least, in this configuration.
We have also considered configurations where the equilateral conditions are not satisfied.

After obtaining the complete predictions
for the quantum trispectrum
in the multi-field DBI inflation model,
we discussed the observational constraints on the non-linearity parameter, for the equilateral configuration
mentioned above. In single field DBI inflation,
some models based on string theory set-ups
suffer  from a severe constraint related with
the geometrical structure of the compact internal spaces.
However, in the presence of the transfer from
the entropy mode, this constraint no longer holds.
Furthermore, since both the tensor-to-scalar ratio
and non-linearity parameter for the bispectrum are suppressed like $T_{\mathcal{RS}}^{-2}$
for a large transfer from the entropy mode to the curvature perturbation,
it is easier to satisfy the corresponding constraints \cite{Langlois:2009ej}. On the other hand,
the trispectrum is enhanced like $\tau_{NL}^{equi} \propto
T_{\mathcal{RS}}^2 f_{NL}^{equi \;2}$ for a given amplitude of the
bispectrum $f_{NL}^{equi}$
, while the global magnitude
of $\tau_{NL}$ is reduced, going like $1/T_{\mathcal{RS}}^2$
in this case.

In order to obtain a concrete value for the
transfer coefficient $T_{\mathcal{RS}}$, we need a specific
model. It would be interesting
to study this mixing in specific
string theory motivated models \cite{KK}.

{\it Note added:} On the day this work appeared in the arXiv,
the paper \cite{RenauxPetel:2009sj} was also submitted
to the arXiv, it also calculates the trispectrum
in multi-field DBI-inflation. While we have concentrated on
the trispectrum coming from the intrinsically quantum four-point functions of the
fields which are generated during inflation,
Ref. \cite{RenauxPetel:2009sj}, by assuming a specific model, showed that the contribution
to the trispectrum coming from the super-horizon non-linear dynamics is also important.

\begin{acknowledgments}
We would like to thank Takahiro Tanaka for interesting
discussions.
SM and FA are supported by JSPS Research Fellowships.
KK is supported by ERC, RCUK and STFC. SM is grateful
to the ICG, Portsmouth for their hospitality when this work
was completed.

\end{acknowledgments}

\appendix

\section{\label{diagramsapp}
The diagrammatic approach to the trispectrum
from the scalar exchange interaction}

The calculation of the integrals in Eq. (\ref{SEintegral})
is rather long and it is easy to forget some terms or permutations.
In this appendix, we present a systematic and simple way to keep track of all the different contributions and permutations to the integrals in (\ref{SEintegral}).
This approach is based on diagrams and simple rules that allows one to read the respective contributions for the scalar exchange four-point function directly from the diagrams.

To calculate the scalar exchange four-point function we need to evaluate
\begin{eqnarray}
&&\langle\Omega| Q_m(0,\mathbf{k_1})Q_n(0,\mathbf{k_2})Q_p(0,\mathbf{k_3})Q_q(0,\mathbf{k_4})|\Omega\rangle^{SE}
\nonumber\\
&&{\hspace{1cm}}= -\int_{-\infty}^0 d\eta\int_{-\infty}^\eta d\tilde\eta \langle
 0|\left[\left[Q_m(0,\mathbf{k_1})Q_n(0,\mathbf{k_2})
Q_p(0,\mathbf{k_3})Q_q(0,\mathbf{k_4}),H_{(3)}^{int}(\eta)\right],H^{int}_{(3)}(\tilde\eta)\right]|0\rangle, \label{SEintegralAppendix}
\end{eqnarray}
where on the l.h.s. of the equation $Q_m$ are the interaction picture fields and $H_{(3)}^{int}$ is the third-order interaction Hamiltonian that can be found in Eq. (\ref{two_field_lead_dbi_hamil_third}) or
\begin{eqnarray}
H^{int}_{(3)} (\eta) &=& H^{int}_{1} (\eta) +
H^{int}_{2} (\eta)+ H^{int}_{3} (\eta)
+ H^{int}_{4} (\eta) + H^{int}_{5} (\eta),
\label{two_field_lead_dbi_hamil_thirdAppendix}
\end{eqnarray}
where the different type of interactions are
\begin{eqnarray}
H_{1} ^{int} (\eta) &\equiv&
\int d^3x g_1  a {Q_\sigma'}^3,\quad
H_{  2} ^{int} (\eta) \equiv
\int d^3x g_2 a Q_\sigma'
(\partial Q_\sigma)^2, \quad
H_{3} ^{int} (\eta) \equiv
\int d^3x g_3 a Q_\sigma' {Q_s'}^2,\nonumber\\
H_{4} ^{int} (\eta) &\equiv&
\int d^3x g_4 a Q_\sigma'
(\partial Q_s)^2,
H_{5} ^{int} (\eta) \equiv
\int d^3x g_5 a Q_s' (\partial Q_\sigma)(\partial Q_s),\label{interactions}
\end{eqnarray}
where here $Q_\sigma$ and
$Q_s$ are the interaction picture fields.
The coupling constants are
\begin{equation}
g_1=M,\quad g_2=-c_s^2M,\quad g_3=M,\quad g_4=c_s^2M,
\quad g_5=-2c_s^2M,
\end{equation}
where $M=-\frac{1}{\sqrt{8X_0 c_s^7}}$.
The integrand of Eq. (\ref{SEintegralAppendix}) can be written as
\begin{eqnarray}
&&\left[\left[Q_m(0,\mathbf{k_1})Q_n(0,\mathbf{k_2})
Q_p(0,\mathbf{k_3})Q_q(0,\mathbf{k_4}),H_{(3)}^{int}(\eta)\right],H^{int}_{(3)}(\tilde\eta)\right]
\nonumber\\
&&= \sum_{i,j=1} ^{5} \biggl\{
F H_{i}^{int} (\eta) H_{j}^{int} (\tilde{\eta})
+ H_{j}^{int} (\tilde{\eta})  H_{i}^{int} (\eta) F
- H_{j}^{int} (\tilde{\eta}) F H_{i}^{int} (\eta)
- H_{i}^{int} (\eta) F  H_{j}^{int} (\tilde{\eta})
\biggr\}
\label{integrand}
\end{eqnarray}
where we have defined $F$ as $F\equiv Q_m(0,\mathbf{k_1})Q_n(0,\mathbf{k_2})Q_p(0,\mathbf{k_3})Q_q(0,\mathbf{k_4})$.

One can immediately see that at the leading order the only non-zero four-point functions are
$\langle\Omega|Q_\sigma(0,\mathbf{k_1})Q_\sigma(0,\mathbf{k_2})Q_\sigma(0,\mathbf{k_3})Q_\sigma(0,\mathbf{k_4})|\Omega\rangle^{SE}$, $\langle\Omega| Q_s(0,\mathbf{k_1})Q_s(0,\mathbf{k_2})Q_\sigma(0,\mathbf{k_3})Q_\sigma(0,\mathbf{k_4})|\Omega\rangle^{SE}$ and $\langle\Omega| Q_s(0,\mathbf{k_1})Q_s(0,\mathbf{k_2})Q_s(0,\mathbf{k_3})Q_s(0,\mathbf{k_4})|\Omega\rangle^{SE}$. The four-point functions
$\langle\Omega| Q_\sigma(0,\mathbf{k_1})Q_\sigma(0,\mathbf{k_2})Q_\sigma(0,\mathbf{k_3})Q_s(0,\mathbf{k_4})|\Omega\rangle^{SE}$
and $\langle\Omega| Q_\sigma(0,\mathbf{k_1})Q_s(0,\mathbf{k_2})Q_s(0,\mathbf{k_3})Q_s(0,\mathbf{k_4})|\Omega\rangle^{SE}$
are zero at this order. This is a direct consequence of the absence of the interactions $Q_s^3$ and $Q_\sigma^2Q_s$ in the third order interaction Hamiltonian (\ref{two_field_lead_dbi_hamil_thirdAppendix}). In terms of diagrams this fact is clear. With the interactions (\ref{interactions}) one cannot draw diagrams with one and three external entropy legs, respectively.

In order to find the non-zero four-point functions we need the following set of rules. We draw the allowed diagrams with two of the vertices (\ref{interactions}), where one leg of the first vertex is glued with one leg of the other vertex (entropy legs can only be glued with entropy legs, similarly for adiabatic legs). The resulting four-point function can be constructed as:
\begin{itemize}
\item Multiply by
$-2(2\pi)^3\delta^{(3)}(\mathbf{K})N^4/(k_1k_2k_3k_4)^{3/2}$,
where $N$ is the normalization of the mode functions.
Multiply by the coupling constants of the vertices that enter the diagram.
The overall minus sign is already present in (\ref{SEintegralAppendix}). The Dirac delta function ensures momentum conservation. The factor $N^4/(k_1k_2k_3k_4)^{3/2}$ comes from the normalization of the four external legs, described by $F$. Finally, the factor of two can be understood if one notes that the contribution to the four-point function coming from the first term in the r.h.s. of Eq. (\ref{integrand}) is real and the operator in the second term is just the Hermitian conjugate of the first term. Similarly with the third and fourth terms in the r.h.s. of Eq. (\ref{integrand}).
\item To each leg of the vertex with a spatial derivative associate the usual factor $-i\mathbf{k}$.
\item The different ways of drawing the same-looking diagram
(ignoring the $k$ labels but with the order
in the left-to-right of the vertices the same)
give a multiplicative symmetry factor.
\item The contribution coming from the third (and fourth) term in the r.h.s. of Eq. (\ref{integrand}) is equal to the contribution coming from the first (and second) term but with the sign of $k_1$ and $k_2$ (but not $k_{12}$), that enter the arguments of $\mathcal{F}_i$ functions, changed to minus sign.

\item In the diagrams with vertices $H_{1} ^{int}$ or $H_{3} ^{int}$
the form factor is made of $\mathcal{F}_1$ functions. The definitions of the $\mathcal{F}_i$ functions can be found in Appendix \ref{Fifc}.
In the diagrams with vertices $H_{2} ^{int}$, $H_{4} ^{int}$ or $H_{5} ^{int}$  the form factor is made of
$\mathcal{F}_2$ functions.
 In the diagrams where the l.h.s. vertex is either one of $H_{1} ^{int}$ or $H_{3} ^{int}$ and the r.h.s. vertex is either one of $H_{2} ^{int}$, $H_{4} ^{int}$ or $H_{5} ^{int}$,
the form factor is made of $\mathcal{F}_3$ functions.
 In the diagrams where the l.h.s. vertex is either one of $H_{2} ^{int}$, $H_{4} ^{int}$ or $H_{5} ^{int}$ and the r.h.s. vertex is either one of $H_{1} ^{int}$ or $H_{3} ^{int}$,
the form factor is made of $\mathcal{F}_4$ functions.
The order of the six arguments in $\mathcal{F}_i$
is read from the diagrams too.
The first three arguments are read from the l.h.s. vertex
and the last three arguments from the r.h.s vertex.
For example, for the diagram made of two $H_{3} ^{int}$ vertices, the first
and fourth arguments of $\mathcal{F}_2$
are the legs with time derivatives.
$\mathbf{k}_1$ in the first vertex and
$\mathbf{k}_4$ in the second vertex.
\item At the end, add twenty three permutations of $\{k_1,k_2,k_3,k_4\}$ to restore the symmetry of the four-point functions under the exchange of the momentum vectors.
\end{itemize}

\begin{figure}[t]
\scalebox{0.8}{
\centerline{
\includegraphics{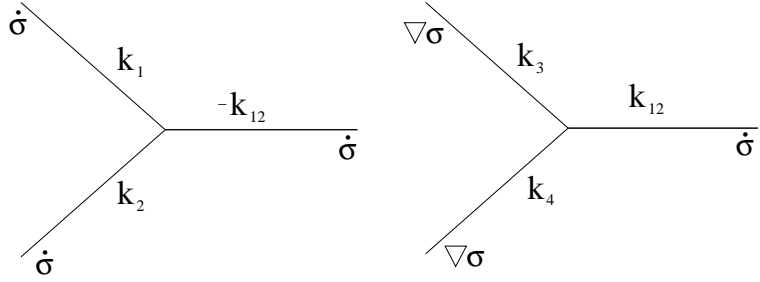}}}
\caption{The two leading order vertices that contribute for the scalar exchange trispectrum in the single field case. Dot over $\sigma$ denotes that in that leg the interaction is a time derivative. $\nabla \sigma$ means that the interaction is a spatial derivative. There is momentum conservation at the vertices, hence at each vertex the momentum vectors have to sum to zero. Overall momentum conservation implies that $\mathbf{k}_{12}=-\mathbf{k}_{34}$, hence the labels in the r.h.s. diagram.}
\label{SingleVertices}
\end{figure}

Let us consider an explicit example and calculate the contribution for the adiabatic four-point function of a diagram where the l.h.s. vertex is $H_{1} ^{int}$ and the r.h.s. vertex is $H_{2} ^{int}$. These vertices are schematically depicted
in Fig. \ref{SingleVertices} and they are the same ones as in the single field case.

\begin{figure}[t]
\scalebox{1}{
\centerline{
\includegraphics{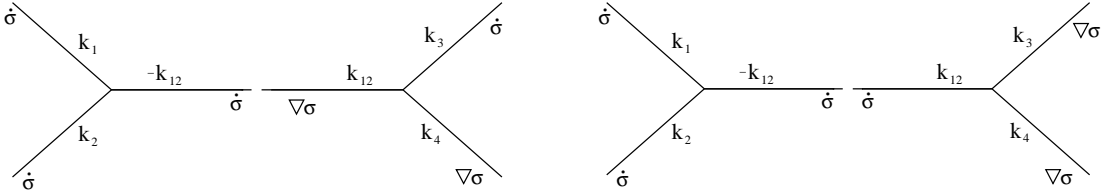}}}
\caption{The two diagrams that contribute to the adiabatic four-point function, where the l.h.s. vertex is $H_{1} ^{int}$ and the r.h.s. vertex is $H_{2} ^{int}$. The first, on the l.h.s. panel, we glue two legs with a time and a spatial derivative. The second, on the r.h.s. panel, we glue two legs with time derivatives.}
\label{SingleDiagrams}
\end{figure}

There are two distinct diagrams that one can draw, as can be seen
in Fig. \ref{SingleDiagrams}, hence there will be two terms.
For the diagram on the l.h.s. of Fig. \ref{SingleDiagrams}
we get a symmetry factor six. This is because there are three ways
we can glue the $\nabla\sigma$ leg to
the $\dot\sigma$ leg and there are two $\nabla\sigma$ legs.
For the diagram on the r.h.s of Fig. \ref{SingleDiagrams}
we get a symmetry factor three. This is because there are three ways
of gluing the $\dot\sigma$ leg of the $H_{2} ^{int}$ vertex with one
of the $\dot\sigma$ legs of the $H_{1} ^{int}$ vertex. In both diagrams,
we have two spatial derivatives so we get a factor
$(-i)^2$. These diagrams will be proportional to
$\mathcal{F}_3$.

Finally, the diagram on the l.h.s. gives
\begin{eqnarray}
&&\!\!\!\!\!\!\!\!\!\!\!\!\!\!\!\!\langle\Omega|Q_\sigma(0,\mathbf{k_1})Q_\sigma(0,\mathbf{k_2})Q_\sigma(0,\mathbf{k_3})Q_\sigma(0,\mathbf{k_4})|\Omega\rangle^{SE}
\nonumber\\
&&\supset(-i)^2 6(2\pi)^3\delta(\mathbf{K})
\frac{-2N^4 g_1g_2}{(k_1k_2k_3k_4)^\frac{3}{2}}
\left(\mathbf{k_{12}}\cdot\mathbf{k_4}\right)
(\mathcal{F}_3(k_1,k_2,-k_{12},k_3,k_4,k_{12})
-\mathcal{F}_3(-k_1,-k_2,-k_{12},k_3,k_4,k_{12}))
\nonumber\\&&
\quad+23\,\mathrm{permutations\, of}\{k_1,k_2,k_3,k_4\},
\end{eqnarray}
and the diagram in the r.h.s. gives
\begin{eqnarray}
&&\!\!\!\!\!\!\!\!\!\!\!\!\!\!\!\!\langle\Omega|Q_\sigma(0,\mathbf{k_1})Q_\sigma(0,\mathbf{k_2})Q_\sigma(0,\mathbf{k_3})Q_\sigma(0,\mathbf{k_4})|\Omega\rangle^{SE}
\nonumber\\
&&\supset
(-i)^2 3 (2\pi)^3\delta(\mathbf{K})
\frac{-2N^4 g_1g_2}{(k_1k_2k_3k_4)^\frac{3}{2}}
\left(\mathbf{k_3}\cdot\mathbf{k_4}\right)
(\mathcal{F}_3(k_1,k_2,-k_{12},k_{12},k_3,k_4)
-\mathcal{F}_3(-k_1,-k_2,-k_{12},k_{12},k_3,k_4))
\nonumber\\&&
\quad+23\,\mathrm{permutations\, of}\{k_1,k_2,k_3,k_4\}
\end{eqnarray}

\begin{figure}[t]
\scalebox{.6}{
\centerline{
\includegraphics{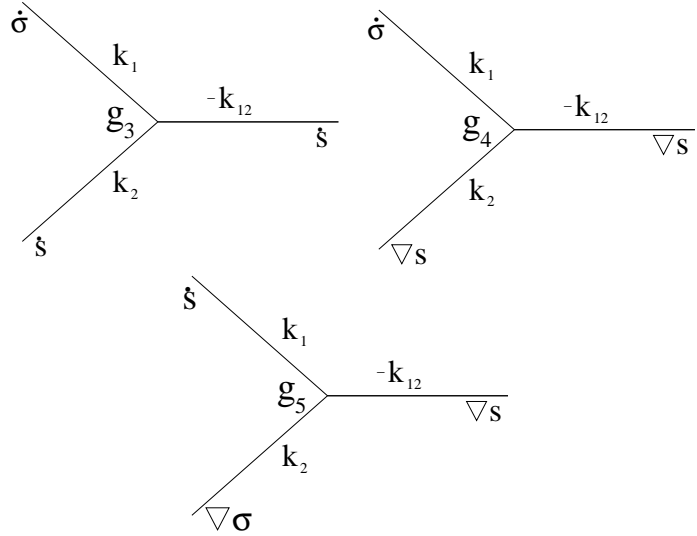}}}
\caption{The leading order vertices containing entropy modes that contribute for the scalar exchange trispectrum in the multi-field case. $g_i$ denotes the coupling constants. As before, the dot over $s$ denotes that in that leg the interaction is a time derivative. $\nabla s$ means that the interaction is a spatial derivative. A label $s$ in the leg means that it is a entropy mode. $\sigma$ denotes an adiabatic leg.}
\label{MultiVertices}
\end{figure}

Now, let us consider another example and calculate the contribution for the mixed four-point function of a diagram where the l.h.s. vertex is $H_{4} ^{int}$ and the r.h.s. vertex is $H_{5} ^{int}$. These diagrams involves vertices that contain entropy legs, they are schematically depicted
in Fig. \ref{MultiVertices}.

Because the entropy and the adiabatic mode functions are
the same, if one substitutes $s$ with $\sigma $
in the vertices of Fig. \ref{MultiVertices} we get the vertices of Fig. \ref{SingleVertices}. Because the $\mathcal{F}_i$ functions only depend
on the mode functions, that are the same
for the entropy and the adiabatic modes we don't need
to define other $\mathcal{F}_i$ functions, and the single field ones are enough.

\begin{figure}[t]
\scalebox{1}{
\centerline{
\includegraphics{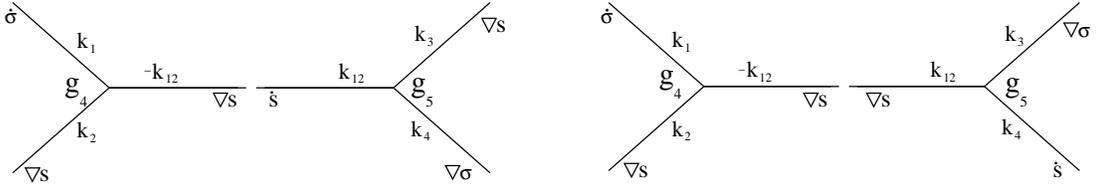}}}
\caption{The two diagrams that contribute to the mixed four-point function, where the l.h.s. vertex is $H_{4} ^{int}$ and the r.h.s. vertex is $H_{5} ^{int}$.}
\label{MultiDiagrams}
\end{figure}

There are two diagrams that contribute in this example, as can be seen
in Fig. \ref{MultiDiagrams}, hence there will be two terms.
In both diagrams, we have four spatial derivatives,
this gives a factor of $(-i)^4$.
The diagram on the l.h.s. of the figure
has a symmetry factor two, because there are two ways of
gluing the $\dot s$ leg of the r.h.s. vertex
with one of the two $\nabla s$ legs of the l.h.s. vertex.
Similarly, the symmetry factor of the r.h.s. diagram is also two.

The l.h.s. diagram contributes with
\begin{eqnarray}
&&\!\!\!\!\!\!\langle\Omega| Q_\sigma(0,\mathbf{k_1})Q_\sigma(0,\mathbf{k_2})Q_s(0,\mathbf{k_3})Q_s(0,\mathbf{k_4})|\Omega\rangle^{SE}+5\,\mathrm{perms.}
\nonumber\\
&&\supset(-i)^4 2
 (2\pi)^3\delta(\mathbf{K})
\frac{-2N^4g_4g_5}{(k_1k_2k_3k_4)^\frac{3}{2}}
\left(\mathbf{k_2} \cdot (-\mathbf{k_{12}})\right)
\left(\mathbf{k_3}\cdot\mathbf{k_4}\right)
(\mathcal{F}_2(k_1,k_2,-k_{12},k_{12},k_3,k_4)
-\mathcal{F}_2(-k_1,-k_2,-k_{12},k_{12},k_3,k_4))
\nonumber\\&&
\quad+23\,\mathrm{permutations\, of}\{k_1,k_2,k_3,k_4\},
\end{eqnarray}
and the diagram in the r.h.s. gives
\begin{eqnarray}
&&\!\!\!\!\!\!\langle\Omega| Q_\sigma(0,\mathbf{k_1})Q_\sigma(0,\mathbf{k_2})Q_s(0,\mathbf{k_3})Q_s(0,\mathbf{k_4})|\Omega\rangle^{SE}+5\,\mathrm{perms.}
\nonumber\\
&&\supset(-i)^4 2 (2\pi)^3\delta(\mathbf{K})
\frac{-2N^4g_4g_5}{(k_1k_2k_3k_4)^\frac{3}{2}}
\left(\mathbf{k_2} \cdot (-\mathbf{k_{12}})\right)
\left(\mathbf{k_{12}} \cdot \mathbf{k_4} \right)
(\mathcal{F}_2(k_1,k_2,-k_{12},k_4,k_3,k_{12})
-\mathcal{F}_2(-k_1,-k_2,-k_{12},k_4,k_3,k_{12}))
\nonumber\\&&
\quad+23\,\mathrm{permutations\, of}\{k_1,k_2,k_3,k_4\},
\end{eqnarray}
where 5 perms. denotes the following permutations of the displayed term $[1\leftrightarrow3]$, $[1\leftrightarrow4]$, $[2\leftrightarrow3]$, $[2\leftrightarrow4]$ and $[1\leftrightarrow3,2\leftrightarrow4]$.

\section{\label{Fifc}Definitions of the $\mathcal{F}_{i}$ functions}

In this Appendix, we present the definitions and the analytical expressions of the four $\mathcal{F}_i$ functions (with $i=1,\ldots,4$) that appear in the main text.

Because there are only four different types of double time integrals over the mode functions, we define four $\mathcal{F}_i$ functions (with $i=1,\ldots,4$). Their explicit expressions are
\begin{eqnarray}
\mathcal{F}_1(k_1,k_2,k_3,k_4,k_5,k_6)&=&\int_{-\infty}^0d\eta a(\eta)\int_{-\infty}^\eta d\tilde\eta a(\tilde\eta)U^{*'}(\eta,k_1)U^{*'}(\eta,k_2)U^{*'}(\eta,k_3)U^{*'}(\tilde\eta,k_4)U^{*'}(\tilde\eta,k_5)U^{*'}(\tilde\eta,k_6)
\nonumber\\
&=&-4\frac{N^6c_s^6}{H^2}|k_1\cdots k_6|^\frac{1}{2}\frac{1}{\mathcal{A}^3\mathcal{C}^3}\left(1+3\frac{\mathcal{A}}{\mathcal{C}}+6\frac{\mathcal{A}^2}{\mathcal{C}^2}\right),
\label{f1_express}
\end{eqnarray}
where $N$ is the normalisation of the mode functions
given by $N=H/\sqrt{2c_s}$, $\mathcal{A}$ is defined by the sum of the last three arguments of the $\mathcal{F}_i$ functions as  $\mathcal{A}=k_4+k_5+k_6$ and $\mathcal{C}$ is defined by the sum of all the arguments as $\mathcal{C}=k_1+k_2+k_3+k_4+k_5+k_6$. The remaining functions are
\begin{eqnarray}
\mathcal{F}_2(k_1,k_2,k_3,k_4,k_5,k_6)&=&\int_{-\infty}^0d\eta a(\eta)\int_{-\infty}^\eta d\tilde\eta a(\tilde\eta)U^{*'}(\eta,k_1)U^{*}(\eta,k_2)U^{*}(\eta,k_3)U^{*'}(\tilde\eta,k_4)U^{*}(\tilde\eta,k_5)U^{*}(\tilde\eta,k_6)
\nonumber\\
&=&-\frac{N^6c_s^2}{H^2}\frac{|k_1k_4|^\frac{1}{2}}{|k_2k_3k_5k_6|^\frac{3}{2}}\frac{1}{\mathcal{A}\mathcal{C}}
\bigg[
      1+\frac{k_5+k_6}{\mathcal{A}}+2\frac{k_5k_6}{\mathcal{A}^2}
      \nonumber\\&&
      +\frac{1}{\mathcal{C}}  \left(k_2+k_3+k_5+k_6+\frac{1}{\mathcal{A}}\left(\left(k_2+k_3\right)\left(k_5+k_6\right)+2k_5k_6\right)+2\frac{k_5k_6\left(k_2+k_3\right)}{\mathcal{A}^2}\right)
      \nonumber\\&&
      +\frac{2}{\mathcal{C}^2}\bigg(k_5k_6+\left(k_2+k_3\right)\left(k_5+k_6\right)+k_2k_3
      \nonumber\\&&
      \qquad\quad+\frac{1}{\mathcal{A}}\left(k_2k_3\left(k_5+k_6\right)+2k_5k_6\left(k_2+k_3\right)\right)+2\frac{k_2k_3k_5k_6}{\mathcal{A}^2}\bigg)
      \nonumber\\&&
      +\frac{6}{\mathcal{C}^3}\left(k_2k_3\left(k_5+k_6\right)+k_5k_6\left(k_2+k_3\right)+2\frac{k_2k_3k_5k_6}{\mathcal{A}}\right)+24\frac{k_2k_3k_5k_6}{\mathcal{C}^4}
\bigg],
\label{f2_express}
\end{eqnarray}
\begin{eqnarray}
\mathcal{F}_3(k_1,k_2,k_3,k_4,k_5,k_6)&=&\int_{-\infty}^0d\eta a(\eta)\int_{-\infty}^\eta d\tilde\eta a(\tilde\eta)U^{*'}(\eta,k_1)U^{*'}(\eta,k_2)U^{*'}(\eta,k_3)U^{*'}(\tilde\eta,k_4)U^{*}(\tilde\eta,k_5)U^{*}(\tilde\eta,k_6)
\nonumber\\
&=&2\frac{N^6c_s^4}{H^2}\frac{|k_1k_2k_3k_4|^\frac{1}{2}}{|k_5k_6|^\frac{3}{2}}\frac{1}{\mathcal{A}\mathcal{C}^3}
\left[
      1+\frac{k_5+k_6}{\mathcal{A}}+2\frac{k_5k_6}{\mathcal{A}^2}
      +\frac{3}{\mathcal{C}}\left(k_5+k_6+2\frac{k_5k_6}{\mathcal{A}}\right)
      +12\frac{k_5k_6}{\mathcal{C}^2}
\right],\nonumber\\
\label{f3_express}
\end{eqnarray}
\begin{eqnarray}
\mathcal{F}_4(k_1,k_2,k_3,k_4,k_5,k_6)&=&\int_{-\infty}^0d\eta a(\eta)\int_{-\infty}^\eta d\tilde\eta a(\tilde\eta)U^{*'}(\eta,k_1)U^{*}(\eta,k_2)U^{*}(\eta,k_3)U^{*'}(\tilde\eta,k_4)U^{*'}(\tilde\eta,k_5)U^{*'}(\tilde\eta,k_6)
\nonumber\\
&=&2\frac{N^6c_s^4}{H^2}\frac{|k_1k_4k_5k_6|^\frac{1}{2}}{|k_2k_3|^\frac{3}{2}}\frac{1}{\mathcal{A}^3\mathcal{C}}
\bigg[1+\frac{\mathcal{A}}{\mathcal{C}}+\frac{\mathcal{A}^2}{\mathcal{C}^2}+\frac{k_2+k_3}{\mathcal{C}}+2\frac{\mathcal{A}\left(k_2+k_3\right)+k_2k_3}{\mathcal{C}^2}
\nonumber\\&&
\qquad\qquad\qquad\qquad\qquad\quad
+3\frac{\mathcal{A}}{\mathcal{C}^3}\left(\mathcal{A}\left(k_2+k_3\right)+2k_2k_3\right)+12k_2k_3\frac{\mathcal{A}^2}{\mathcal{C}^4}\bigg],
\label{f4_express}
\end{eqnarray}
where the function $U(\eta,k)$ is a modified mode function given by
\begin{equation}
U(\eta,k)=N\frac{1}{|k|^\frac{3}{2}}\left(1+ikc_s\eta\right)e^{-ikc_s\eta}.
\end{equation}
If the sign of the argument $k$ of $U$ is positive then $U$ is equal to the mode function, if the sign is negative then $U$ equals the complex conjugate of the mode function.

\section{\label{TDBI}The scalar exchange trispectrum in the equilateral limit}

In this Appendix, we present the scalar exchange trispectrum of the primordial curvature perturbation
$\mathcal{R}$  in the equilateral configuration.
These analytical expressions can be derived from the general expression for the scalar exchange trispectrum,
Eq.~(\ref{4point_curv_scalar_exchange}).
We will use these simpler equations
to study the shape dependence of the trispectrum
and to calculate the non-linearity parameter $\tau_{NL}$
in subsection  \ref{subsec:Shape}.

At leading order in the small sound speed and in the slow-roll expansion, the scalar exchange trispectrum
is
\begin{eqnarray}
{\langle\Omega|\mathcal{R}(0,\mathbf{k}_1)
\mathcal{R}(0,\mathbf{k}_2)\mathcal{R}(0,\mathbf{k}_3)\mathcal{R}(0,\mathbf{k}_4)|\Omega\rangle^{SE}}&=&
{\langle\Omega|\mathcal{R}(0,\mathbf{k}_1)
\mathcal{R}(0,\mathbf{k}_2)\mathcal{R}(0,\mathbf{k}_3)\mathcal{R}(0,\mathbf{k}_4)|\Omega\rangle^{SE}}_{4 \sigma}
\nonumber\\
&&
+
{\langle\Omega|\mathcal{R}(0,\mathbf{k}_1)
\mathcal{R}(0,\mathbf{k}_2)\mathcal{R}(0,\mathbf{k}_3)\mathcal{R}(0,\mathbf{k}_4)|\Omega\rangle^{SE}}_{\sigma \sigma ss}
\nonumber\\
&&+
{\langle\Omega|\mathcal{R}(0,\mathcal{R}{k}_1)
\mathcal{R}(0,\mathbf{k}_2)\mathcal{R}(0,\mathbf{k}_3)\mathcal{R}(0,\mathbf{k}_4)|\Omega\rangle^{SE}}_{4s},
\end{eqnarray}
where the three different contributions come from the terms in Eq.~(\ref{4point_curv_scalar_exchange})
that are coming from the purely adiabatic, mixed
and purely entropy components.

Explicitly, they are given by
\begin{eqnarray}
{\langle\Omega|\mathcal{R}(0,\mathbf{k}_1)
\mathcal{R}(0,\mathbf{k}_2)\mathcal{R}(0,\mathbf{k}_3)\mathcal{R}(0,\mathbf{k}_4)|\Omega\rangle^{SE}}_{4\sigma}
&=&(2\pi)^3\delta^{(3)}(\mathbf{K})
\frac{1}{c_s^4}
\frac{1}{(1+T_{\mathcal{R}\mathcal{S}}^2)^3}\mathcal{D}_1(k_{12}/k)\tilde {\cal P}_\mathcal{R*}^2(k)\tilde {\cal P}_\mathcal{R*}(k_{12}) +2\,\mathrm{perms},\nonumber\\
\end{eqnarray}
\begin{eqnarray}
{\langle\Omega|\mathcal{R}(0,\mathbf{k}_1)
\mathcal{R}(0,\mathbf{k}_2)\mathcal{R}(0,\mathbf{k}_3)\mathcal{R}(0,\mathbf{k}_4)|\Omega\rangle^{SE}}
_{\sigma \sigma ss}&=& (2\pi)^3\delta^{(3)}(\mathbf{K})
\frac{T_{\mathcal{R} \mathcal{S}}^2}
{(1+T_{\mathcal{R}\mathcal{S}}^2)^3}
 \frac {1}{c_s^4}
\mathcal{D}_2(k_{12}/k)
\tilde {\cal P}_\mathcal{R*}^2(k)
\tilde {\cal P}_\mathcal{R*}(k_{12})
+2\,\mathrm{perms},\nonumber\\
\end{eqnarray}
\begin{eqnarray}
{\langle\Omega|\mathcal{R}(0,\mathbf{k}_1)
\mathcal{R}(0,\mathbf{k}_2)\mathcal{R}(0,\mathbf{k}_3)\mathcal{R}(0,\mathbf{k}_4)|\Omega\rangle^{SE}}_{4s}
&=&(2\pi)^3\delta^{(3)}(\mathbf{K})
\frac{T_{\mathcal{R} \mathcal{S}}^4}
{(1+T_{\mathcal{R}\mathcal{S}}^2)^3}
\frac{1}{c_s^4}
\mathcal{D}_3(k_{12}/k)
\tilde {\cal P}_\mathcal{R*}^2(k)\tilde {\cal P}_\mathcal{R*}(k_{12}) +2\,\mathrm{perms},\nonumber\\
\end{eqnarray}
where we have defined three $\mathcal{D}_i(x)$ functions, with $i=1,2,3$, as
\begin{eqnarray}
\mathcal{D}_1(x)&=&\frac{x^4(61952 + 48384 x - 10240 x^2 - 20832 x^3 - 5184 x^4 + 1412 x^5 +
 824 x^6 + 103 x^7)}{2^9 (x+2)^4},\nonumber\\
\mathcal{D}_2(x)&=&\frac{x^4
(31232+23744x-6272x^2-10496x^3-1248x^4+1904x^5+824x^6+103x^7)}
{2^8 (x+2)^4},\nonumber\\
\mathcal{D}_3(x)&=&\frac{x^4
(512 - 896 x - 2304 x^2 - 160 x^3 + 2688 x^4 + 2396 x^5 +
   824 x^6 + 103 x^7)}{2^9 (x+2)^4},
\end{eqnarray}
where $k_{12}=k\sqrt{2(1+\cos\theta_3)}$,
$k_{13}=k\sqrt{2(1+\cos\theta_2)}$ and
$k_{14}=k\sqrt{2(1+\cos\theta_1)}$. $k$ denotes the common amplitude of
all four momentum vectors $\mathbf{k}_j$, with $j=1,\ldots,4$ and the
rescaled power spectrum is $\tilde {\cal
P}_\mathcal{R*}(k)=H^2/(4\epsilon c_sk^3)$ that should be evaluated at
sound horizon crossing.

It is worth mentioning that from
Eqs.~(\ref{CI_simple_rel})
and (\ref{SE_simple_rel}), the following relation holds,
\begin{eqnarray}
\mathcal{D}_2 = \mathcal{D}_1+\mathcal{D}_3.
\end{eqnarray}


\end{document}